\documentclass[12pt]{article}
  
\usepackage{amsmath,amssymb}

\usepackage{xspace}  
\usepackage{epsfig}  
  
\usepackage{graphicx}               % Standard graphics package  

\newcommand{\nc}{\newcommand}  

\nc{\beq}{\begin{equation}}  
\nc{\eeq}{\end{equation}}  
\nc{\beqa}{\begin{eqnarray}}  
\nc{\eeqa}{\end{eqnarray}}  
\nc{\bea}{\begin{eqnarray}}  
\nc{\eea}{\end{eqnarray}}  
\nc{\ra}{\rightarrow}  
\nc{\lsim}{\begin{array}{c}\,\sim\vspace{-21pt}\\< \end{array}}  
\nc{\gsim}{\begin{array}{c}\sim\vspace{-21pt}\\> \end{array}}

\nc{\LL}{L}  
\nc{\vv}{\tilde{v}}  
\nc{\kk}{\tilde{k}}  
\nc{\GG}{\widetilde{G}}  
\nc{\MM}{\ensuremath{\mathcal{M}}}  
\nc{\UU}{\ensuremath{\mathcal{U}}}  
\nc{\ZZ}{\ensuremath{\mathcal{{\cal{Z}}}}}  
\nc{\MQ}{\ensuremath{\hat{M}_{Q}}}
\nc{\Mu}{\ensuremath{\hat{M}_{u}}}
\nc{\Md}{\ensuremath{\hat{M}_{d}}}
\nc{\Mtu}{\ensuremath{\tilde{M}_{u}}}
\nc{\Mtd}{\ensuremath{\tilde{M}_{d}}}

\title{  
\vspace*{-2.3cm}  
\begin{flushright}  
\normalsize{  
ANL-HEP-PR-06-53\\
CU-TP-1152\\ 
EFI-06-08\\  
FERMILAB-PUB-06/234-T\\
hep-ph/0607106  
  }  
\end{flushright}  
\vspace{1.5cm}  
\Large  
\textbf{
Light Kaluza-Klein States
in Randall-Sundrum Models with
  Custodial $SU(2)$}\vspace*{1.0cm}   
\author{\large\textbf{Marcela Carena$^a$},  
\textbf{Eduardo Pont\'{o}n$^{b}$},
\\[0.3cm]  
\textbf{Jos\'{e} Santiago$^{a}$},  
and  
\textbf{C.E.M.~Wagner$^{c,d}$}\\ \\[0.5cm]  
$^a$\normalsize\emph{Fermi National Accelerator Laboratory,  
P.O. Box 500, Batavia, IL 60510, USA} \\  
$^b$\normalsize\emph{Department of Physics, Columbia University,}\\
\normalsize\emph{538 W. 120th St, New York, NY 10027, USA} \\  
$^c$\normalsize\emph{HEP Division, Argonne National Laboratory,  
9700 Cass Ave.,  
Argonne, IL 60439, USA} \\  
$^d$\normalsize\emph{Enrico Fermi Institute and Kavli Institute for
  Cosmological Physics,} \\ 
\normalsize\emph{Univ. of Chicago, 5640  
Ellis Ave., Chicago, IL 60637, USA}}
\date{}}  
\begin{document}  
\setcounter{page}{0}  
\maketitle  
%\date{}  
%\vspace*{1cm}  
\begin{abstract}  
We consider Randall-Sundrum scenarios based on $SU(2)_{L} \times
SU(2)_{R}$ and a discrete parity exchanging $L$ with $R$.  The
custodial and parity symmetries can be used to make the tree level
contribution to the $T$ parameter and the anomalous couplings of the
bottom quark to the $Z$ very small.  We show that the resulting
quantum numbers typically induce a negative $T$ parameter at one loop
that, together with the positive value of the $S$ parameter, restrict
considerably these models.  There are nevertheless regions of
parameter space that successfully reproduce the fit to electroweak
precision observables with  light Kaluza-Klein excitations
accessible at colliders.  We consider models of gauge-Higgs
unification that implement the custodial and parity symmetries 
and find that the electroweak data singles
out a very well defined region in parameter space.
In this
region one typically finds light gauge boson Kaluza-Klein
excitations as well as light $SU(2)_{L}$ singlet, and sometimes
also doublet, fermionic states, that mix with the top quark, and that
may yield interesting signatures at future colliders.
\end{abstract}  
  
\thispagestyle{empty}  
\newpage  
  
\setcounter{page}{1}

\baselineskip18pt

%-----------------------------------------------------------------------------
\section{Introduction}
\label{sec:intro}  
%-----------------------------------------------------------------------------

The LHC era that is about to start is expected to unravel the
mysteries of electroweak (EW) symmetry breaking and the origin of the gauge
hierarchy.  Among the theories that attempt to explain the hierarchy
problem, the Randall-Sundrum model~\cite{Randall:1999ee} stands out as
one of the most promising ones.  The well-known little hierarchy
problem becomes, however, critical in such a model.  In fact, it is
not just a matter of fine-tuning, but also that the
EW precision data turn out to be so constraining that the masses of
the Kaluza-Klein (KK) modes of bulk fields are pushed to scales that
make their discovery at the LHC extremely challenging.
Even though placing the fermions (together with the gauge
bosons) in the bulk softens the constraints, there is still a very
large contribution to the Peskin-Takeuchi~\cite{Peskin:1991sw} $T$
parameter~\cite{Huber:2000fh}.  The presence of infrared (IR) brane
kinetic terms~\cite{Carena:2002dz} can reduce the size of the oblique
corrections and, if large enough, induce a spectrum of gauge KK modes
accessible at colliders~\cite{Carena:2003fx,Carena:2004zn}.  In this
paper we will consider a different alternative based on custodial
symmetry.  It was noticed in~\cite{Agashe:2003zs} that extending the
bulk gauge symmetry to a \textit{custodially symmetric} $SU(2)_L
\times SU(2)_R$, also reduces the tree level contribution to the $T$
parameter to phenomenologically allowed levels for masses of the KK
excitations of the gauge bosons $M_{KK}\sim 3$ TeV. At the same time, 
the Right-Handed
(RH) quarks were included in doublets under the $SU(2)_R$ symmetry bringing
with them $SU(2)_R$-symmetric partners that in the case of the RH top
quark can be very light.  Unfortunately, 
this mode mixes with the bottom quark and
induces anomalous couplings to the $Z$ 
that again puts strong constraints on these models, pushing the KK modes
beyond experimental detectability.  However, it has been very recently
pointed out~\cite{Agashe:2006at} that the custodial symmetry
\textit{together} with a discrete $L\leftrightarrow R$ symmetry can
protect the $b$ coupling to the $Z$ from anomalous corrections,
therefore apparently saving the last hurdle for a realistic
Randall-Sundrum model with small brane kinetic terms and 
observable KK modes at the LHC. 
In this article we show that, although the custodial and $LR$ parity
symmetries are very powerful in rendering the 
tree level contributions to the $T$ parameter and the 
$Z \bar{b}_L b_L$ coupling very
small, the quantum numbers we are forced upon to get such a protection
-- bidoublets under the $SU(2)_L \times SU(2)_R$ symmetry -- typically
induce sizable and negative contributions to $T$ at the loop level. 
Such negative 
contribution to the $T$ parameter, together with a positive value of
the $S$ parameter greatly constraint the masses of the KK modes.  A
scan over parameter space shows however that there are regions where
the sign of $T$ is reversed and gauge boson KK masses, $M_{KK}\sim 3$ TeV, 
can be compatible with experimental data.  
Hence, we obtain the first
Randall-Sundrum model with negligible brane kinetic terms that is fully
compatible with EW precision observables and has gauge boson
KK masses accessible at the LHC.

The above scenarios are based on a fundamental Higgs field, and suffer
from the little hierarchy problem.
An alternative theory
of EW symmetry breaking, that does not require any
fundamental scalars, has received a lot of attention recently in the
context of models with extra dimensions.  These are the models of
``gauge-Higgs unification''~\cite{gauge:Higgs:unification,Agashe:2004rs}, 
where the Higgs field is a
pseudo-Nambu-Goldstone boson that arises as the component along the
extra dimensions of gauge fields of broken symmetries.  The higher
dimensional gauge symmetry protects the Higgs from cut-off sensitive
corrections making its potential, that arises at the quantum level,
finite and therefore calculable.  A very simple, yet realistic example
is based on an $SO(5)\times U(1)_X$ bulk symmetry broken to $SO(4)
\times U(1)_X\sim SU(2)_L\times SU(2)_R \times U(1)_X$ on the IR brane
and to $SU(2)_L\times U(1)_Y$ on the UV brane.  The Higgs field
corresponds to the zero mode of the $A_5$ gauge boson along the broken
direction $SO(5)/SO(4)$~\cite{Agashe:2004rs}.  The extended gauge 
symmetry makes
models of gauge-Higgs unification very predictive and therefore also
very constrained. In particular, a light Higgs is a generic prediction
of these models. Also, the fact that Yukawa couplings are really gauge
couplings reduces the freedom to play with the location of the zero
modes and makes it more difficult to cancel the negative contribution
to the $T$ parameter generated by the bidoublets.  Interestingly, the
allowed regions of parameter space typically lead to fermion states
well under a TeV.

The structure of the paper is the following.  In section~\ref{sec:LRmodels}
we introduce the custodially symmetric version of the RS model with a
fundamental Higgs, not necessarily localized on the IR brane, 
and review the constraints coming from oblique
parameters and the $Z\bar{b}b$ coupling.  We show that, 
as already noted in Ref.~\cite{Agashe:2006at}, in the absence of
large brane kinetic terms, in order
to obtain light enough KK states, accessible at the LHC, the
Left-Handed (LH)
quarks are required to belong to bidoublets of $SU(2)_L  \times SU(2)_R$.
In section~\ref{sec:Tparameter} we compute in
detail the one-loop contributions to the $T$ parameter and show that
the bidoublets induce a sizable, negative $T$ in most regions of
parameter space.  We then consider models of gauge-Higgs unification
in section~\ref{sec:gaugHiggs}, and identify the regions of parameter
space allowed by the EW precision measurements.  In
section~\ref{phenomenology} we present the most relevant features of
the phenomenology of this type of models,
and we
conclude in section~\ref{sec:conclusion}.

%-----------------------------------------------------------------------------
\section{$SU(2)_{L} \times SU(2)_{R}$ scenarios}
\label{sec:LRmodels}
%-----------------------------------------------------------------------------

We consider an $SU(2)_{L} \times SU(2)_{R} \times U(1)_{X}$ gauge 
theory on a slice of $AdS_5$ with metric 
\beqa
ds^{2} = e^{-2ky} \eta_{\mu\nu} dx^{\mu} dx^{\nu} -dy^{2},
\eeqa
and fifth-dimensional coordinate $0\leq y \leq L$.
The fermions are allowed to propagate in the
bulk.  In order to address the gauge hierarchy
problem, the Higgs field has to be localized near the IR brane ($y=L$).
We will analyze both the case of 
a Higgs exactly localized on the IR brane, as well as the case of a Higgs
propagating in the bulk, as it occurs in gauge-Higgs unification 
scenarios.

The gauge group is broken by boundary conditions to the Standard Model (SM)
on the UV brane ($y=0$).  
This is done with the following assignment of
boundary conditions
\bea
W^a_{L\,\mu} \sim (+,+)~, \quad B_\mu \sim (+,+)~, \\
W^b_{R\,\mu} \sim (-,+)~, \quad Z^\prime_\mu \sim (-,+)~, 
\eea
where $+$ ($-$) stands for Neumann (Dirichlet) boundary conditions, 
$a=1,2,3$, $b=1,2$ and $B_\mu$ and $Z^\prime_\mu$ are the
following two combinations of neutral gauge bosons
\bea
B_\mu=\frac{g_{5\,X} W^3_{R\, \mu}
+g_{5\,R} X_\mu}{\sqrt{g^2_{5\,R}+g^2_{5\,X}}}~, \quad \quad \quad
Z^\prime_\mu=\frac{g_{5\,R} W^3_{R\, \mu}
-g_{5\,X} X_\mu}{\sqrt{g^2_{5\,R}+g^2_{5\,X}}}~, 
\eea
with $g_{5\,R}, g_{5\,X}$ the five-dimensional coupling constants of
the $SU(2)_{R}$ and $U(1)_{X}$ groups, respectively.
The covariant derivative in the basis of well defined parities then
reads
\begin{equation}
D_\mu=\partial_\mu - \mathrm{i} \left [
g_{5\,L} W^a_{L\,\mu} T^a_L+ 
g_{5}^\prime \frac{Y}{2} B_\mu +
g_{5\,R} W^b_{R\,\mu} T^b_R+ 
g_{5\,Z^\prime} Q_{Z}^\prime Z^\prime_\mu 
\right]~,
\end{equation}
where the hypercharge and $Z^\prime$ gauge couplings are
\beq
g_5^\prime=\frac{g_{5\,R}\,g_{5\,X}}{\sqrt{g^2_{5\,R}+g^2_{5\,X}}}~,
\quad \quad \quad
g_{5\,Z^\prime}=\sqrt{g^2_{5\,R}+g^2_{5\,X}}~,
\eeq 
whereas the charges are
\begin{equation}
\frac{Y}{2}=T^3_R+Q_X~, \quad \quad \quad Q_Z^\prime=\frac{g_{5\,R}^2 
T^3_R - g_{5\,X}^2 Q_X}
{g_{5\,R}^2+g_{5\,X}^2}~,
\end{equation} 
so that the electric charge reads
\beqa
Q =T^3_L+ T^{3}_{R} + Q_X~.
\label{charge}
\eeqa
The modes with $(+,+)$ boundary conditions have zero modes that make
the gauge bosons of the SM whereas the ones with
$(-+)$ boundary conditions only have massive modes.  We can now
integrate out these and the KK excitations of the SM gauge bosons to
obtain a four-dimensional effective theory that can be compared to
experiment.

%-----------------------------------------------------------------------------
\subsection{Tree-level Corrections from Gauge KK Modes}
\label{sec:gaugecorrections}
%-----------------------------------------------------------------------------

The massive gauge bosons induce tree-level corrections to the SM 
gauge boson masses and to their couplings to the fermions (as
well as four-fermion interactions).  Given that the KK scale is well
above the EW breaking scale, we may treat the Higgs vacuum
expectation value (vev)
perturbatively and keep only the leading corrections.  These can be
expressed in terms of the zero-momentum gauge boson propagators for
the massive KK modes obeying $(+,+)$ boundary conditions. More precisely,
in terms of the coefficient of 
$P_{\mu\nu} = \eta_{\mu\nu} - p_{\mu} p_{\nu}/p^2$,
with zero mode parts subtracted~\cite{Carena:2003fx} :
\beqa
\GG^{++}_{p=0}(y,y') &=& \frac{1}{4k(kL)} \bigg\{ \frac{1-e^{2k L}}{kL} +
e^{2ky_{<}} (1-2ky_{<}) + e^{2ky_{>}} \left[ 1+ 2k (L-y_{>})\right]
\bigg\}~,
\label{Gpp}
\eeqa
and those obeying $(-,+)$ boundary conditions:
\beqa
\GG^{-+}_{p=0}(y,y') &=& - \frac{1}{2k} \left[ e^{2ky_{<}} - 1 \right]~.
\label{Gmp}
\eeqa
Here $p$ is the 4-dimensional momentum and  $y_{<}$
($y_{>}$) denote the smallest (largest) of $y$ and $y'$, the
fifth-dimensional coordinate.

To leading order in the corrections, the SM gauge boson
masses are
\beqa
m^{2}_{Z} &=& \frac{e^{2} v^{2}}{2s^{2} c^{2}} \bigg\{ 1 + 
\frac{e^{2}}{s^{2} c^{2}} \left[
\delta^{2}_{++} + \left(\frac{g_R^2}{g_L^2}c^{2} - s^{2}
\right) \delta^{2}_{-+}
\right]
 + \cdots \bigg\}~,
\\ [0.5em]
m^{2}_{W} &=& \frac{e^{2} v^{2}}{2s^{2}} \bigg\{ 1 + \frac{e^{2}}{s^{2}} 
\left[
\delta^{2}_{++} +\frac{g_R^2}{g_L^2} \delta^{2}_{-+}
\right]
 + \cdots \bigg\}~,
\eeqa
where $v = 174~\rm{GeV}$ is the SM Higgs vev, 
four-dimensional couplings are defined in terms of the
five-dimensional ones as $g_L=g_{5\,L}/\sqrt{L}$ (similarly for the
rest), and we have defined as it is customary
\begin{equation}
  e=\frac{g^\prime\,g_L}{\sqrt{g_L^2+g^{\prime\, 2}}}~, \quad
  s=\frac{e}{g_L}~,  
\quad c=\sqrt{1-s^2}~.
\end{equation}
The corrections from exchange of the towers of the
$W^{a}_{L}$ and hypercharge $B$ gauge bosons
are contained in
\beqa
\delta^{2}_{++} &=& \frac{L v^{2}}{2} \int^{L}_{0} \! dy dy' e^{-2ky} 
f_{H}(y)^{2}
\GG^{++}_{p=0}(y,y') e^{-2ky'} f_{H}(y')^{2}~.
\label{v2pp}
\eeqa
The function
$f_{H}(y)$ is the Higgs
profile which is kept arbitrary for the time 
being.  The
contributions from the $W^{b}_R$ and the neutral $Z'$ towers are encoded in
$\delta^{2}_{-+}$, which is defined as in Eq.~(\ref{v2pp}), but in
terms of the massive propagator given in Eq.~(\ref{Gmp}).

The corrections to the couplings of the SM gauge bosons to the fermion
currents depend on the fermion zero-mode wavefunctions.  The couplings
to the $Z$ take the form
\beqa
\label{Zcouplings}
\hspace*{-3cm} 
 \frac{e}{s c} Z_{\mu} \sum_{\psi} \bigg\{ \bar{\psi}^{(0,r)} 
\gamma^{\mu} \left( T^{3}_{L} - s^{2} Q \right) \psi ^{(0,r')}
\left[ \delta^{rr'} + \frac{e^{2}}{s^{2}c^{2}} G^{rr'}_{++} \right] 
\nonumber \\ [0.5em]
\mbox{} - \frac{e^{2}}{s^{2}c^{2}} \bar{\psi}^{(0,r)} \gamma^{\mu} 
\left( \frac{g_R^2}{g_L^2}c^{2} T^{3}_{R} + s^{2} T^{3}_{L} - s^{2} Q 
\right) \psi ^{(0,r')} 
G^{rr'}_{-+}
\bigg\}~,
\hspace*{-3cm} 
\eeqa
where the sum runs over all chiral fermions. Here,
\beqa
G^{rr'}_{++} &=& \frac{v^{2}}{2} \int^{L}_{0} \! dy dy' 
\left[ f^{0}_{\psi^{r}}(y) \right]^{*} f^{0}_{\psi^{r'}}(y)
\GG^{++}_{p=0}(y,y') e^{-2ky'} f_{H}(y')^{2}~,
\label{Gfpp}
\eeqa
where $\GG^{++}_{p=0}(y,y')$ was defined in Eq.~(\ref{Gpp}), and
$f^{0}_{\psi^{r}}(y)$ is the zero-mode wavefunction for the fermion
$\psi^{r}$, $r$ being a flavor index.  $G^{rr'}_{-+}$ is defined
analogously in terms of the propagator of Eq.~(\ref{Gmp}).

Similarly, the charged currents read
\beqa
\label{Wcouplings}
\hspace*{-3cm} 
\frac{e}{\sqrt{2}s} W^{+}_{\mu} \sum_{\psi} \bigg\{ 
\bar{\psi}^{(0,r)} \gamma^{\mu} T^{+}_{L} \psi^{(0,r')}
\left[ \delta^{rr'} + \frac{e^{2}}{s^{2}} G^{rr'}_{++} \right] 
+ \frac{g_R^2}{g_L^2}\frac{e^{2}}{s^{2}} \bar{\psi}^{(0,r)} \gamma^{\mu} 
T^{+}_{R} \psi^{(0,r')} G^{rr'}_{-+}
\bigg\} + {\rm h.c.} 
\hspace*{-3cm} 
\eeqa
where $T_{L,R}^+ = T_{L,R}^1 + i T_{L,R}^2$ and
the sum runs over all chiral fermions.

In the general case, the previous corrections are flavor-dependent and
would set very stringent bounds on the masses of the KK modes.
However, it is well known that these effects can be controlled
effectively in RS scenarios by localizing the fermions of the first
two generations closer to the UV brane than to the IR brane 
\cite{Huber:2002sp}.  In fact,
in this region of parameter space the quantities $G^{rr'}_{++}$ and
$G^{rr'}_{-+}$ become essentially independent of the fermion profile,
thus leading to universal effects that can be recast in the form
of oblique corrections.  Furthermore, in this same region 
it becomes possible to generate the fermion mass hierarchies
entirely from the overlaps between the Higgs and fermion
wavefunctions.  Therefore, we concentrate on this scenario, which
allows us to parametrize the corrections to the Z-pole observables
measured at LEP and SLD together with the mass of the $W$ measured at
the Tevatron and LEP2 in terms of the Peskin-Takeuchi $S$, $T$ and $U$
parameters \cite{Peskin:1991sw}.

For the first two quark and lepton generations
as well as the third generation leptons, the general couplings to the $Z$ and
$W$ given in Eqs.~(\ref{Zcouplings}) and (\ref{Wcouplings}) reduce to
\beqa
\label{Zqq}
\frac{e}{s c} Z_{\mu} \left[ \bar{\psi}^{(0,r)} 
\gamma^{\mu} \left( T^{3}_{L} - s^{2} Q \right) \psi^{(0,r)} \right]
\left( 1 + \frac{e^{2}}{s^{2}c^{2}} \, G^{f}_{++} \right)~,
\eeqa
and
\beqa
\label{Wqq}
\frac{e}{\sqrt{2} s} W_{\mu}^+ \left[ \bar{\psi}^{(0,r)}
\gamma^{\mu} T^{+}_{L} \psi^{(0,r)} \right]
\left( 1 + \frac{e^{2}}{s^{2}} \, G^{f}_{++} \right)~,
\eeqa
where we denote by $G^f_{++}$ the common value of Eq.~(\ref{Gfpp}) for
the fermions with $c_{i}>1/2$, and used the fact that the
corrections proportional to $G^{rr'}_{-+}$ in Eqs.~(\ref{Zcouplings})
and (\ref{Wcouplings}) become negligible for such values of 
$c_{i}$.  Since
these shifts are universal they may be absorbed by a rescaling of the
gauge boson fields which allows to describe these effects in terms of
the oblique parameters $S$, $T$ and $U$.  In fact, to describe the
Z-pole observables it is possible to take into account the most
important non-oblique corrections, coming from KK exchange
contributions to the Fermi constant, $G_{F}$, by using the effective
parameters of \cite{Carena:2002dz}
\begin{eqnarray}
S_{\rm eff} &=& 32 \pi G^f_{++}~,\nonumber \\
T_{\rm eff} &=& \frac{8 \pi}{c^2} G^f_{++} -
\frac{4 \pi}{c^2} \left[ \delta_{++}^2 - \delta_{-+}^2 \right]
 + \frac{2 \pi v^2}{s^2} G^{\mu\mu}_{++}~,
\label{STU}
\\
U_{\rm eff} &=& -8 \pi v^2 G^{\mu\mu}_{++}~.
\nonumber
\end{eqnarray}
In the above,
\beqa
G^{\mu\mu}_{++} &=& \frac{1}{L} \int^{L}_{0} \! dy dy' 
|f_\mu^{(0)}(y)|^2 
\GG^{++}_{p=0}(y,y') 
|f_\mu^{(0)}(y^\prime)|^2~
\label{Gfmumu}
\eeqa
represents the contribution to muon decay from the exchange of the KK
towers of the $SU(2)_{L}$ gauge bosons,
with $f_\mu^{(0)}(y)$ the wave function of the muon zero mode. The
Higgs vev is $v = 174~{\rm GeV}$.
In general, there are also terms proportional to 
$G^f_{-+}$ and $G^{\mu\mu}_{-+}$,
that have not been included in Eq.~(\ref{STU}) 
since they become negligibly small when
the leptons and the first two quark generations are localized close to the 
UV brane.  Furthermore, as is apparent from Eq.~(\ref{Zcouplings}), 
these corrections depend on the relative values of $T^3_R$ and $T^3_L$. 
Hence, whenever
these corrections become important, non-universal fermion-gauge-boson
couplings are induced, even when all fermions are localized identically. 
Thus, although localizing the leptons and first two quark generations 
near the conformal point
can be interesting due to the possibility of a small coupling to the
lightest gauge boson KK modes,
%obtaining $S=0$, and 
a determination of the bounds requires a global
fit to the EW observables and will be presented
elsewhere~\cite{gaugeHiggs}. In this paper we will
concentrate on a region of parameter space where an analysis based on the 
effective parameters $S$, $T$ and $U$ is sufficient.
Consequently, we have not included such terms in
Eqs.~(\ref{STU}).

We see in the expression for $T$ in Eq.~(\ref{STU}) the custodial
symmetry mechanism at work: the largest contribution coming from the
fermion-independent terms parametrized by $\delta^2_{++}$ of
Eq.~(\ref{v2pp}) is partially canceled by the corresponding
contribution associated with the $Z^\prime$ gauge boson,
parametrized by $\delta^{2}_{-+}$.  The cancellation is not perfect
due to the breaking of the custodial symmetry 
induced by the choice of boundary conditions.  
Thus, one finds that the tree-level contribution to the $T$ parameter
is not identically zero,
although, in practice, it is quite small.
We have evaluated the effects due
to the KK-tower contributions to $G_{F}$, which are also
rather small in the region we are considering, so that $U_{\rm eff}$
is very small.  Thus, in Eq.~(\ref{STU}) it is the $S$ parameter that
receives the largest contributions, and generically leads to strong
bounds on the present class of scenarios.  It is important to
note that in the models under study, in which the light
fermions are localized far from the IR brane, the
$S$ parameter is always positive.
In section~\ref{sec:Tparameter}
we will show that
the residual loop-level contributions to the $T$ parameter are also
quite relevant.

The third quark generation doublet needs to be localized closer to the IR
brane in order to generate the large top mass.  As a result, one
should introduce an additional parameter that allows to describe the
difference in the corrections to the coupling of the bottom quark with
respect to the light quarks.  The accurate measurement of $R_b$, the ratio
of the width of the Z-boson decay into bottom-quarks and the total
hadronic width, puts strong constraints on this difference. Since the
LH bottom coupling to the Z is roughly five times larger
than the RH one, this mostly constrains the former. 
The RH coupling, in turn, is mainly constrained
by the measurement of the forward-backward and left-right bottom
quark asymmetries measured at the LEP and SLD colliders. However, 
these constraints are much weaker than those affecting the LH
coupling~\cite{Haber:1999zh,Beauty}. Moreover, since the experimental
value of $R_b$ is approximately one standard deviation larger than 
the value predicted by the SM, positive and negative 
corrections on this coupling are not equally constrained. Approximately,
at the 2-$\sigma$ level, assuming no large correction to the RH
bottom coupling, the constraint on the correction to the 
LH bottom coupling to the $Z$ reads~\cite{Beauty},
\begin{equation}
 -2 \times 10^{-3} \lsim 
\frac{\delta g_{b\,L}}{g_{b\,L}} 
\lsim 6 \times 10^{-3}.
\label{epsilonb-bound}
\end{equation}
Since the LH bottom couples to the $SU(2)_{R}$ and/or $U(1)_X$ gauge bosons,
the shift in its coupling to the $Z$, given in Eq.~(\ref{Zcouplings}),
receives a contribution from the terms proportional to
$G^{b_{L}}_{-+}$.
After rescaling the $Z$ wave function so that its couplings to
the first two generations do not present anomalous shifts, the value
of the anomalous LH bottom quark coupling to the $Z$ is given by
\begin{equation}
\frac{\delta g_{b\,L}}{g_{b\,L}} 
= \frac{e^2}{s^2 c^2} 
\left[
G^{b_{L}}_{++} - 
\frac{c^2 T^3_R g_R^2/g_L^2+s^2 T^3_L-s^2 Q}{T^3_L -s^2 Q}
G^{b_{L}}_{-+} - 
G^f_{++} \right] ~,
\label{epsilonb}
\end{equation} 
where $T^3_L = -1/2$ and $Q = -1/3$ are the $SU(2)_L$ isospin and
charge quantum numbers of the LH bottom quark, respectively.  
$T^3_{R}$ is the bottom quark $SU(2)_R$ isospin, which is determined
by the embedding into the 5D gauge group representation.

For the case of $T_R^3 = 0$, negative corrections to 
$\delta g_{b\,L}/g_{b\,L}$ are generated, that become 
larger in absolute value as the LH bottom is localized closer to 
the infrared brane. Indeed, for $T^3_R(b_L)=0$,
Eq.~(\ref{epsilonb}) gives
\begin{equation}
\frac{\delta g_{b\,L}}{g_{b\,L}}(T^3_R=0) 
= \frac{e^2}{s^2 c^2} 
\left[
G^{b_{L}}_{++} - 
s^2 \frac{T^3_L-Q}{T^3_L -s^2 Q}
G^{b_{L}}_{-+} - 
G^f_{++} \right] 
\approx \frac{e^2}{s^2 c^2} 
\left[
G^{b_{L}}_{++} - 
0.09 
G^{b_{L}}_{-+} - 
G^f_{++} \right] ~.
\label{epsilonb:dobletes}
\end{equation} 
These corrections are depicted by the curve
labeled by ``gauge'' in Fig.~1, which we will explain in more detail
below, and become small as we separate $b_L$ (and $t_L$) from the IR
brane, 
in which case the coupling of the bottom
and the light fermions to the $Z$ become similar. 
Note however that $t_L$ cannot be too far from the IR brane, since this
would lead to an unacceptably small top quark Yukawa coupling.

We can also see how, when the LH bottom acquires a
non-vanishing value of $T^3_R$, the
$SU(2)_R$ custodial symmetry can protect the bottom coupling to the
$Z$~\cite{Agashe:2006at}.  If we set $T^3_L=T^3_R$ for $b_L$, together
with $g_R=g_L$, the anomalous coupling greatly simplifies
\begin{equation}
\frac{\delta g_{b\,L}}{g_{b\,L}}(T^3_R=T^3_L, g_R=g_L) 
= \frac{e^2}{s^2 c^2} 
\left[
G^{b_{L}}_{++} - 
G^{b_{L}}_{-+} - 
G^f_{++} \right] ~,
\label{epsilonb:bidobletes}
\end{equation} 
and the contributions from $G^{b_{L}}_{++}$ and $G^{b_{L}}_{-+}$ tend
to cancel as a result of the custodial symmetry and the quantum
numbers of the bottom.  As was
the case for the $T$ parameter, the breaking due to the boundary
conditions makes the cancellation imperfect.  Notice that the degree
of cancellation depends on both the bottom localization and the Higgs
profile.  
Furthermore, since $G^{b_{L}}_{++}$ and $G^{b_{L}}_{-+}$ are
different even when the bottom and Higgs fields are exactly localized
on the IR brane, a residual effect still remains in this
limit.\footnote{When both the Higgs and bottom   
fields are exactly
localized on the IR brane, one finds 
$(G^{b_{L}}_{++} -G^{b_{L}}_{-+})/G^{b_{L}}_{++} \approx -1/kL$.  
When the Higgs is not
exactly localized, the residual difference is about twice as large due
to the tail of the Higgs wavefunction that ``feels'' the breaking of
the custodial symmetry away from the IR brane.} 
In fact, one finds that the difference
$G^{b_{L}}_{++} - G^{b_{L}}_{-+}$ increases slightly as the LH bottom
comes closer to the IR brane.

Since $\delta g_{b\,L}$ measures the deviation of the
bottom coupling to the $Z$ relative to the couplings of the first two
generations, Eq.~(\ref{epsilonb:bidobletes}) receives an extra
contribution from $G^{f}_{++}$.
In particular, for an appropriate localization of the
fermion fields, it is possible for the terms in
Eq.~(\ref{epsilonb:bidobletes}) to cancel exactly, but we do not
explore such a possibility here.  Thus, although the anomalous
$Z\bar{b}_Lb_L$ coupling is small, it is not necessarily irrelevant.

%-----------------------------------------------------------------------------
\subsection{Tree-level Corrections from Fermion KK Modes}
\label{sec:epsilonb}
%-----------------------------------------------------------------------------

In addition to the corrections due to the presence of massive spin-1
fields given in the previous section there are extra tree-level
contributions to the couplings of the fermions to the SM gauge fields.
These arise from integrating out the fermion KK modes that mix with
the fermion zero-modes through the Higgs vev.  Given that we are
taking the first two generations to be localized away from the Higgs
field, for the light quarks and leptons these mixing effects are
exponentially suppressed due to the associated zero-mode wavefunction.
However, for the third generation such effects can be important, as
pointed out in \cite{Agashe:2005dk}.

As we will see, in the models of Ref.~\cite{Agashe:2003zs} this
constraint by itself is strong enough to put the higher dimensional
physics beyond the reach of the LHC. We conclude that a mechanism that
suppresses the anomalous $Z\bar{b}_Lb_L$ coupling as outlined at the end
of the previous section seems to be an essential ingredient for these
models to be viable.

Thus, we start by studying the class of models with custodial $SU(2)$,
where the SM LH top and bottom arise from 5D $SU(2)_{R}$
singlets and the RH top arises from an $SU(2)_{R}$ doublet:
\beqa
\begin{array}{rclcrcl}
q_{L} &=& \begin{pmatrix} 
q^{t}_{L}(+,+) \\
q^{b}_{L}(+,+)
\end{pmatrix}~,
& \hspace{5mm} &
Q_{R} &=& \begin{pmatrix} 
t_{R}(+,+) \\
b'_{R}(-,+)
\end{pmatrix}~,
\end{array}
\eeqa
where, under $SU(2)_{L} \times SU(2)_{R}$, $q \sim (2,1)$ and $Q \sim
(1,2)$.  We also indicated the boundary conditions by assigning
parities at the UV and IR branes, respectively.\footnote{As for gauge
bosons, a $-$ parity assignment stands for a Dirichlet boundary
condition at the corresponding brane.  For fermions, a Dirichlet
boundary condition for a given 4D chirality fixes, through the
equations of motion, the boundary condition obeyed by the opposite
chirality.  We denote this boundary condition by a $+$ parity
assignment.  It is, in general, a condition on the first derivative
with respect to the extra-dimensional coordinate.} The choice of
parities is determined by the requirement that the low-energy theory
should have a LH doublet and a RH top, and by the
requirement that $SU(2)_{R}$ be preserved on the IR brane.  We exhibit
only the parities for the chiralities containing zero modes.  The
parity assignments for the opposite chiralities can be simply read
from these.

The important point is that reproducing the top mass requires the
localization of the zero mode in $Q$ near the IR brane.  (As we
mentioned in the previous section and will make explicit below, the LH
doublet $q_L$ cannot be taken too close to the IR brane due to large
corrections to the anomalous $b$ couplings.)  This in turn implies
that the lightest mode of $b'_{R}$, a state with the quantum numbers
of the RH bottom, becomes rather light.  Its tree-level
mixing with the LH bottom induces large anomalous couplings
of the latter to the $Z$ gauge boson.

\begin{figure}[t]  
\centerline{\includegraphics[width=0.8\textwidth]{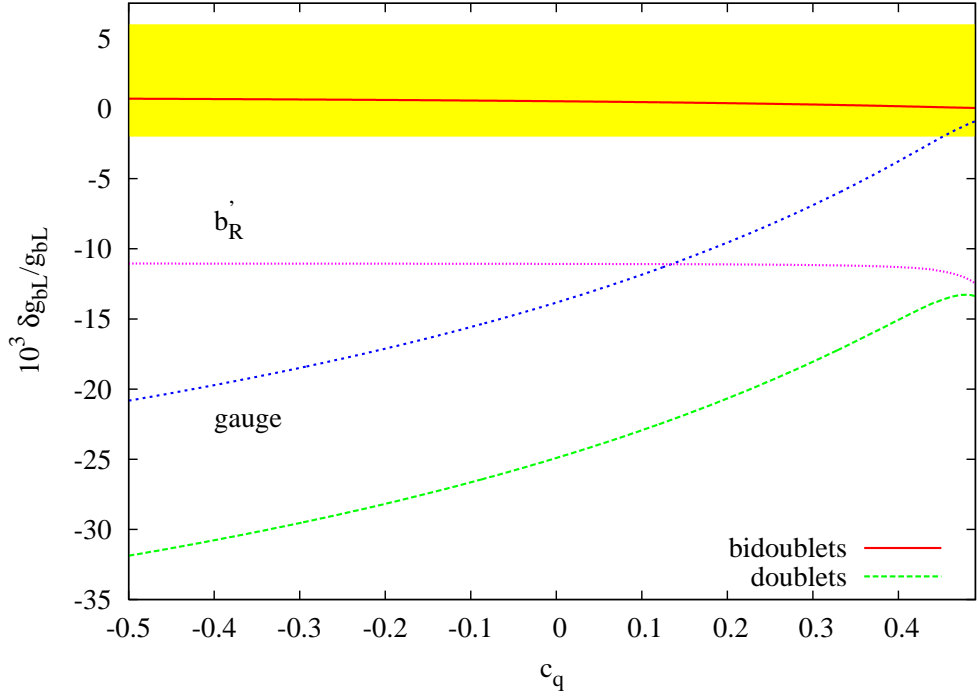}}
\caption{Contribution to $\delta g_{b\,L}/g_{b\,L}$ from
Eq.~(\ref{epsilonb:dobletes}), labeled by ``gauge'', the contribution
due mixing with the lightest modes of $b'_{R}$, and the sum of the
two, for models based on $SU(2)_{R}$ doublets and
for $\kk = k e^{-kL} = 1.5$~TeV.  The
solid line gives $\delta g_{b\,L}/g_{b\,L}$ from Eq.~(\ref{epsilonb})
for models based on bidoublets of $SU(2)_{L} \times SU(2)_{R}$.  It is
assumed that the Higgs is localized on the IR brane.  The band
is the current $2\sigma$ bound.}
\label{fig:epsilonb}  
\end{figure}  
In Fig.~\ref{fig:epsilonb} we plot the minimum obtainable $\delta
g_{b\,L}/g_{b\,L}$
as a function of $c_{q}$, the localization parameter of the
$SU(2)_{L}$ doublet, assuming the Higgs is exactly localized on the IR
brane.  We include the contribution due to exchange of KK gauge bosons
in the case that the LH bottom is a singlet of $SU(2)_{R}$
and $g_L=g_R$, given by Eq.
(\ref{epsilonb:dobletes}),
as well as the contribution due to the mixing with the lightest
$b'_{R}$ states.  The gauge contribution depends only on $c_{q}$ and
the KK scale $\tilde{k} = k\,e^{-kL}$.  The contribution due to mixing
depends on $c_{Q}$, the localization parameter for
$t_{R}$, and only very slightly on $c_q$. In fact,  
since the top mass is given in terms of the zero-mode
wavefunctions by
\beqa
m_{\rm top} &=& \lambda_{4} \, f_{q^{(0)}_{L}}(L) f_{t^{(0)}_{R}}(L) \, v~,
\eeqa
where $\lambda_{4} = \lambda_{5}/L$ and $\lambda_{5}$ is the 5D top
Yukawa coupling, we can express the mixing between $b^{(0)}_{L}$ and
$b'^{(1)}_{R}$ as
\beqa
\frac{\lambda_{4} v^{2}}{m^{2}_{b'^{(1)}_{R}}} 
f_{q^{(0)}_{L}}(L) f_{b'^{(1)}_{R}}(L)
&=& \frac{m_{\rm top} v}{m^{2}_{b'^{(1)}_{R}}} 
\frac{f_{b'^{(1)}_{R}}(L)}{f_{t^{(0)}_{R}}(L)}~,
\eeqa
which depends only on $c_{Q}$.  The $c_{Q}$ dependence enters both
through the wavefunctions and the mass of the lightest $b'_{R}$ state.
The absolute value of this function reaches a minimum for 
$c_{Q} \approx  0$.

The curve marked as $b'_{R}$ in Fig.~\ref{fig:epsilonb} corresponds to
the choice of $c_{Q}$ that minimizes the contribution to
$\delta g_{b\,L}/g_{b\,L}$, and is almost constant as a function of
$c_{q}$ (the slight $c_{q}$ dependence is due to the higher KK modes).
We took $\tilde{k} = 1.5~{\rm TeV}$, that corresponds to gauge boson
masses of approximately $3.75~{\rm TeV}$.  We thus see that $\delta
g_{b\,L}/g_{b\,L}$
gives a rather model independent bound on scenarios where the
RH top arises from a 5D $SU(2)_{R}$ doublet, and that the
bounds are rather severe, most likely putting the 
gauge boson KK sates beyond the
reach of upcoming collider experiments.  For example, even for $c_{Q}
\approx 0.5$, the constraint Eq.~(\ref{epsilonb-bound}) leads to $\kk
\gsim 3.5~{\rm TeV}$, or gauge boson KK masses starting at $8.75~{\rm
TeV}$.

In Fig.~\ref{fig:epsilonb}, we also plot $\delta g_{b\,L}/g_{b\,L}$ 
in models where the SM doublets arise from 5D 
$SU(2)_{L} \times SU(2)_{R}$ bidoublets, 
for which the condition $T^3_R=T^3_L$ can be
fulfilled for $b_L$.  In this case there are either no states with the
quantum numbers of the LH bottom in the multiplets that
couple through the top Yukawa coupling, or they come in pairs whose
effects cancel against each other (due to the symmetry exchanging $L$
with $R$).  There can be states that mix with the LH bottom
and give a non-vanishing contribution to $\delta g_{b\,L}/g_{b\,L}$,
coming from the multiplet that gives rise to the RH bottom.
However, these contributions can be easily suppressed through the
bottom quark Yukawa coupling.  Therefore, we only plot the
contribution due to the exchange of gauge KK states, as given in
Eq.~(\ref{epsilonb:bidobletes}).  The anomalous couplings are easily
within the experimental limits, which gives a strong motivation for
including the mechanism of Ref.~\cite{Agashe:2006at} to suppress
contributions to $\delta g_{b\,L}/g_{b\,L}$.  In the next sections we
concentrate on further constraints on scenarios with 
$SU(2)_{L} \times SU(2)_{R}$ bidoublets.

%-----------------------------------------------------------------------------
\section{$SU(2)_{L} \times SU(2)_{R}$ bidoublets and the $T$ parameter}
\label{sec:Tparameter}
%-----------------------------------------------------------------------------

The custodial $SU(2)$ symmetry, which is broken only by boundary
conditions, ensures that the contribution to the T parameter due to
fermion loops is finite.  Nevertheless, the finite contributions
involving the KK modes of the top quark impose significant constraints.
In fact, we will see that making the top-bottom doublet part of
a bidoublet of $SU(2)_L \times SU(2)_R$   
implies a rather definite prediction for the
1-loop contributions to the T parameter.  We concentrate here on the
top sector since it gives the largest effects. Other one-loop contributions
to T, due to gauge bosons and light fermions, are much 
smaller~\cite{Agashe:2003zs}. 

The cancellation of the contributions to $\delta g_{b\,L}/g_{b\,L}$
advocated in \cite{Agashe:2006at} requires that the LH bottom
be part of a $SU(2)_{L} \times SU(2)_{R}$ bidoublet, with $T^3_{L} =
T^{3}_{R}$, i.e. $T^3_{L} + T^{3}_{R} = - 1$, which implies from
Eq.~(\ref{charge}) that the bidoublet $U(1)_{X}$ charge is fixed to
$X=2/3$.  Therefore, in addition to the top partner, the LH
bottom is accompanied by a charge $2/3$ field, $\chi^d$, and a charge
$5/3$ field, $\chi^{u}$.  Writing the Higgs field as a bidoublet of
$SU(2)_{L} \times SU(2)_{R}$, we see that to write a top Yukawa
coupling, the top $SU(2)_{L}$ singlet must arise either from an
$SU(2)_{R}$ singlet or an $SU(2)_{R}$ triplet, with charge $X = 2/3$
[the bottom $SU(2)_{L}$ singlet then must come from an $SU(2)_{R}$
triplet with charge $X = 2/3$].  In the latter case, the $P_{LR}$
parity that ensures the cancellation of $\delta g_{b\,L}/g_{b\,L}$
requires an additional $SU(2)_{L}$ triplet, with couplings related to
those of the $SU(2)_{R}$ triplet.  Thus, we have the following
possible assignments:
\beqa
\begin{array}{rclrcl}
Q_{L} &=& \begin{pmatrix} 
\chi^{u}_{L}(-,+) & q^{t}_{L}(+,+) \\
\chi^{d}_{L}(-,+) & q^{b}_{L}(+,+)
\end{pmatrix}~,
& \hspace{2mm} &
t_{R} (+,+)
\end{array}~,
\label{Bidoubletsinglet}
\eeqa
or
\beqa
\begin{array}{rclrclrcl}
Q_{L} &=& \begin{pmatrix} 
\chi^{u}_{L}(-,+) & q^{t}_{L}(+,+) \\
\chi^{d}_{L}(-,+) & q^{b}_{L}(+,+)
\end{pmatrix}~,
& \hspace{1mm} &
T_{1R} =
\begin{pmatrix} 
\psi^{\prime}_{R}(-,+) \\ 
t^{\prime}_{R}(-,+) \\
b^{ \prime}_{R}(-,+)
\end{pmatrix}~, 
& \hspace{1mm} &
T_{2R} =
\begin{pmatrix} 
\psi^{\prime \prime}_{R}(-,+) \\ 
t_{R}(+,+) \\
b^{\prime \prime}_{R}(-,+)
\end{pmatrix} 
\end{array}~,
\label{Bidoublettriplet}
\eeqa
where, in the bidoublets, $SU(2)_{L}$ acts vertically and $SU(2)_{R}$
acts horizontally.  $T_{1}$ and $T_{2}$ transform as $(3,1)$ and
$(1,3)$, respectively, under $SU(2)_{L} \times SU(2)_{R}$.  In
Eq.~(\ref{Bidoublettriplet}), the parity assignments on $Q$ and
$T_{2}$ are determined by the requirement that $SU(2)_{R}$ be
preserved on the IR brane, together with the zero-mode spectrum.  The
parity of the state $b^{\prime}$ in $T_{1}$ is chosen to be identical
to that of $b^{\prime \prime}$ in $T_{2}$ to avoid large contributions
to $\delta g_{b\,L}/g_{b\,L}$ from mixing of the bottom with light
bottom-like states.

We are interested in the contributions to the $T$ parameter arising
from 1-loop diagrams involving the KK modes in the above cases.  We
concentrate on the assignments shown in Eq.~(\ref{Bidoubletsinglet}),
but at the end of the section we comment on the results in the case
that the top quark arises from the multiplets in
Eq.~(\ref{Bidoublettriplet}).  There are also contributions due to the
remaining multiplets needed to embed the SM, but these can
be easily suppressed either by taking moderately small 5D Yukawa
couplings in those sectors or by choosing their localization
parameters such that the overlap of their zero-mode with the Higgs is
exponentially suppressed. The latter occurs when the light fermion
modes are localized close to the UV brane, as we have assumed in this
work. In the top sector, however, accommodating
the top mass imposes strong constraints on such contributions.

Since the sums over KK mode loops are finite due to the nonlocal
breaking of the custodial $SU(2)$ symmetry, it is possible to compute
$T$ by including only the lowest lying states.  We compute it by
numerically diagonalizing the mass matrix of KK modes, including the
mixings induced by the Higgs vev, and calculating the self-energies
for $W^1_{\mu}$ and $W^3_{\mu}$, via \cite{Lavoura:1992np}
\beqa
T &=&\frac{N_{c}}{16\pi s^{2} c^{2} m^{2}_{Z}} \times
\nonumber \\
& &\left\{\sum_{\alpha} \sum_{i}
\left( V^{L}_{\alpha i} V^{L*}_{\alpha i} + V^{R}_{\alpha i} 
V^{R*}_{\alpha i} \right)
\theta_{+}({\cal M}^{\alpha \alpha}_{u}, {\cal M}^{i i}) +
2 \,{\rm Re}\left( V^{L}_{\alpha i} V^{R\ast}_{\alpha i} \right)
\theta_{-}({\cal M}^{\alpha \alpha}_{u}, {\cal M}^{i i}) \right.
\\ & & \left. \hspace{5mm} \mbox{} -
\sum_{\alpha} \sum^{\alpha - 1}_{\beta}
\left( U^{L}_{\alpha \beta} U^{L*}_{\alpha \beta} + 
U^{R}_{\alpha \beta} U^{R*}_{\alpha \beta} \right)
\theta_{+}({\cal M}^{\alpha \alpha}_{u}, {\cal M}^{\beta \beta}_{u}) +
2 \, {\rm Re}\left( U^{L}_{\alpha \beta} U^{R\ast}_{\alpha \beta} \right)
\theta_{-}({\cal M}^{\alpha \alpha}_{u}, {\cal M}^{\beta \beta}_{u})
\right\}~,
\nonumber 
\label{ExactT}
\eeqa
where $N_{c} = 3$ is the number of colors
and
\beqa
\theta_{+}(y_{1},y_{2}) &=& y^{2}_{1} + y^{2}_{2} - 
\frac{2 y^{2}_{1} y^{2}_{2}}{y^{2}_{1} - y^{2}_{2}} 
\ln \frac{y^{2}_{1}}{y^{2}_{2}}~,
\\
\theta_{-}(y_{1},y_{2}) &=& 2 y_{1} y_{2} 
\left( \frac{2 y^{2}_{1} y^{2}_{2}}{y^{2}_{1} - 
y^{2}_{2}} \ln \frac{y^{2}_{1}}{y^{2}_{2}} - 2 \right)~.
\eeqa
${\cal M}_{u}$ is the diagonal matrix of charge $2/3$ states ($q^{t}$,
$\chi^{d}$ and $t$), and ${\cal M}$ contains the (diagonal) masses for the
remaining states that do not mix among themselves ($q^{b}$ and
$\chi^{u}$).  $V^{L}$ ($V^{R}$) is the matrix of couplings of
LH (RH) fermion fields to $W^{1}_{\mu}$
in the mass eigenstate basis, and
$U^{L}$ ($U^{R}$) is the corresponding matrix of couplings of the
charge $2/3$ states to $W^{3}_{\mu}$.  The matrices 
$U^{L,R}$ are hermitian and satisfy the relations
\beqa
(U^{L,R})^{2} &=& V^{L,R} V^{L,R\dagger}~, \nonumber \\
U^{L} {\cal M}_{u} U^{R} &=& V^{L} {\cal M} V^{R\dagger}~, \\
{\cal M} &=& V^{L\dagger} {\cal M}_{u} V^{R}~, \nonumber 
\eeqa
which can be used to show that the UV divergences associated with the
4D momentum integration cancel in the expression for $T$.

To obtain the contribution to the $T$ parameter due to the new
physics, we need to subtract the SM model top quark contribution
\beqa
 T_{\rm top} &=& \frac{N_c m^{2}_{\rm top}}{16 \pi s^2 c^2 m^{2}_{Z}}~,
 \label{Ttop}
 \eeqa
where $m_{\rm top}$ is the would-be zero-mode mass, obtained after
diagonalization of the mass matrix.  We have checked that the result
converges fast with the number of KK modes.

\begin{figure}[t]  
\centerline{\includegraphics[width=0.8\textwidth]{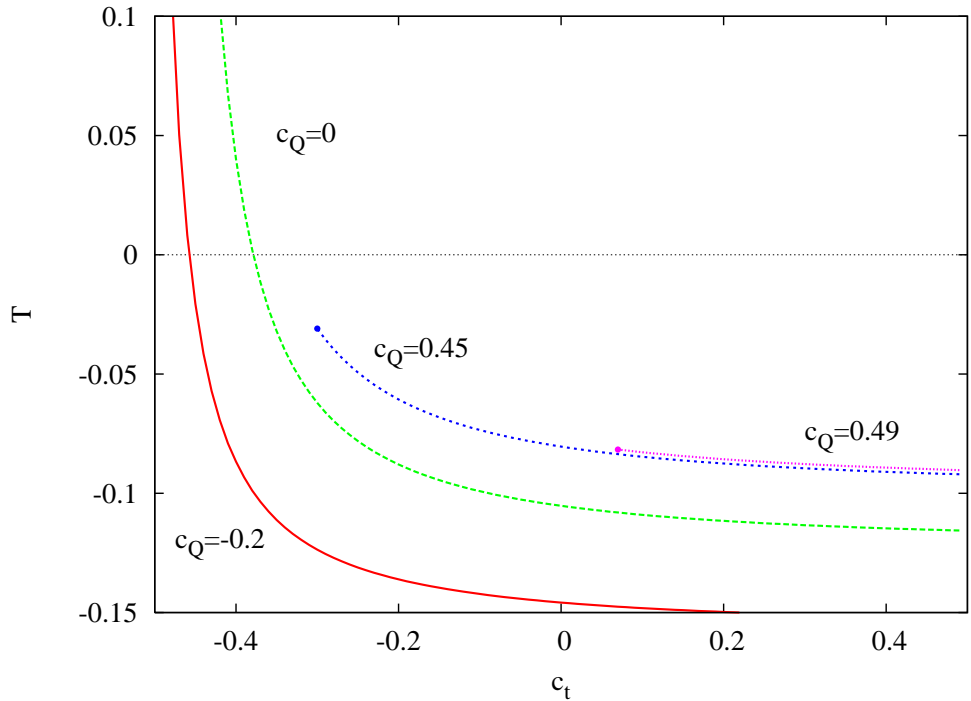}}
\caption{Contribution to the $T$ parameter involving the KK modes of
Eq.~(\ref{Bidoubletsinglet}), which couple to the Higgs through the
top Yukawa coupling.  We use $\tilde{k} = 1.5~{\rm TeV}$ and $m_{\rm
top} = 167~{\rm GeV}$.  The dots indicate the point beyond which the
theory is strongly coupled at the scale of the first KK mode.  It is
assumed that the Higgs field is exactly localized on the IR brane.}
\label{fig:TbidoubletsLocal}  
\end{figure}  
\begin{figure}[t]  
\centerline{\includegraphics[width=0.8\textwidth]{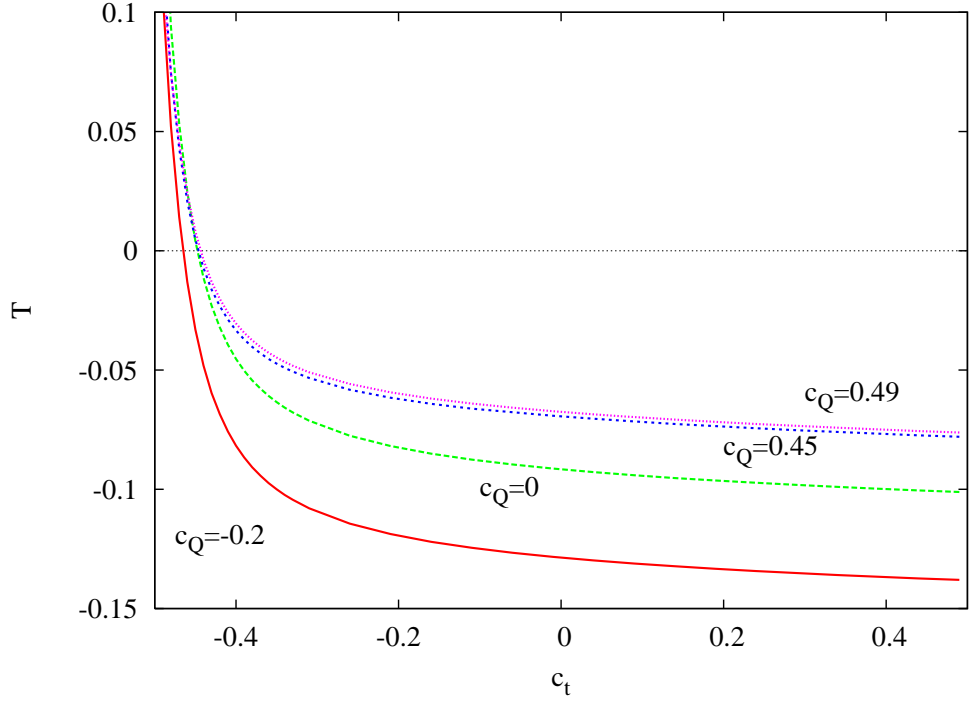}}
\caption{Contribution to the $T$ parameter involving the KK modes of
Eq.~(\ref{Bidoubletsinglet}), which couple to the Higgs through the
top Yukawa coupling.  We use $\tilde{k} = 1.5~{\rm TeV}$ and $m_{\rm
top} = 167~{\rm GeV}$.  It is assumed that the Higgs field has the profile
of gauge-Higgs unification models, Eq.~(\ref{higgs:norm}).}
\label{fig:TbidoubletsDelocal}  
\end{figure}  
In Fig.~\ref{fig:TbidoubletsLocal} we show the $T$ parameter arising
from the multiplets in Eq.~(\ref{Bidoubletsinglet}), with the top
quark contribution subtracted, as a function of $c_{t}$ and for
several values of $c_{Q}$, the localization parameters of the singlet
and bidoublet, respectively.  It is assumed that the Higgs is
localized on the IR brane.  Fig.~\ref{fig:TbidoubletsDelocal} shows
the case where the Higgs is ``maximally'' 
delocalized~\cite{Davoudiasl:2005uu}, as would be the
case in scenarios of gauge-Higgs unification (see
section~\ref{sec:gaugHiggs} below).  There is also a contribution to
$T$ coming from the gauge sector, as given in Eq.~(\ref{STU}), which
is not included in Figs.~\ref{fig:TbidoubletsLocal} or
\ref{fig:TbidoubletsDelocal}.  It is independent of $c_{Q}$ and
$c_{t}$, and is subdominant.

The figures exhibit the following features:
\begin{itemize}

\item $T$ becomes more negative as $c_{Q}$ decreases, which localizes
the bidoublet zero-modes near the IR brane.

\item As $c_{t}$ increases, which localizes $t_{R}$ near the IR brane,
$T$ becomes negative.

\item If we separate $t_{R}$ sufficiently from the IR brane, $T$ can
become positive.  However, in doing so one is forced to increase the
5D top Yukawa coupling to reproduce the top mass, eventually entering the
strong coupling regime, 
i.e. the one-loop corrections are of the same order 
as the tree-level coupling.  We have cut the curves when the theory is strongly
coupled at the scale of the first KK mode.  
Thus, depending on the
localization of the bidoublet, $c_{Q}$, $T$ may never reach positive
values.

\end{itemize}

In fact, the negative contribution to the $T$ parameter is a direct
consequence of the embedding of the SM $SU(2)_{L}$
doublets into bidoublets of $SU(2)_{L} \times SU(2)_{R}$.  To clarify
this point we derive approximate expressions in the case where $t$
is a singlet of $SU(2)_{R}$, as in Eq.~(\ref{Bidoubletsinglet}).
Given that the mixing masses are small compared to the scale of KK
masses, it is justified to treat them perturbatively.  Also, since the
KK sums are convergent, we may keep only the first KK modes of the
bidoublet and singlet.  For clarity we consider separately the case
with only the vector-like bidoublet KK modes, and the
case with only the vector-like singlet KK modes  included.  
The general (and
more complicated) case can be understood from these simplified
results.

\begin{figure}[t]  
\centerline{\includegraphics[width=0.8\textwidth]{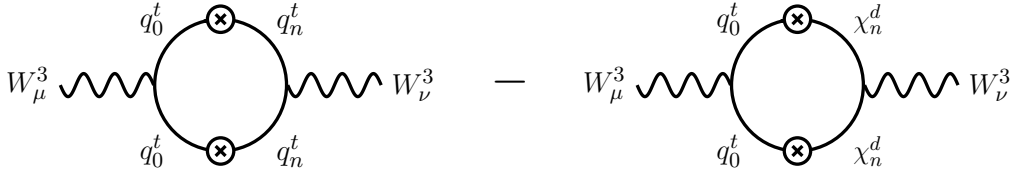}}
\caption{Diagrams that give the dominant contribution to the $T$
parameter when only vector-like bidoublets are present.  When the
$SU(2)_{L}$ singlet top quark is also an $SU(2)_{R}$ singlet as in
Eq.~(\ref{Bidoubletsinglet}), there are no diagrams contributing to
$\Pi_{11}$.  We show explicitly the relative minus sign between the
isospin charges of $q^{t}$ and $\chi^{d}$.  The top mass insertions in
the zero-mode propagator are resummed to all orders.  Each cross
represents an insertion of the EW breaking mass mixing the singlet $t$
with $q^{t}_{n}$ or $\chi^{d}_{n}$.}
\label{fig:diagramsTbidoublet}  
\end{figure}

Let us first consider the case that includes, in addition to the SM
top and bottom, a single vector-like $SU(2)_{L} \times SU(2)_{R}$
bidoublet (in the scenarios at hand, the first KK excitations of the
5D bidoublet fields).  We allow for the KK mass of the $\chi$'s to be
different from the mass of the $q$'s, since they obey different
boundary conditions.  However, since we are treating the EW breaking
vev perturbatively, we use common KK masses $M_{\chi}$ and $M_{q}$ for
their upper and lower $SU(2)_{L}$ components.  There are also EW
breaking masses that mix the (zero-mode) singlet $t$ with the
bidoublet components $q^t$ and $\chi^{d}$.  These mixing masses, which
we call $m_{q^{t},t}$ and $m_{\chi^{d},t}$ respectively, depend on the
integral over the extra dimension of the Higgs zero-mode profile and
the fermion wavefunctions (for the zero-mode of $t$ and the first KK
modes of $q^{t}$ or $\chi^{d}$).  With this notation, the lowest order
contribution to the $T$ parameter takes the form
\beqa
\Delta T = T_{\rm top} \frac{m^{2}_{\chi^{d},t}}{M^{2}_{q}} \left[ 
- F_{1}\left( \frac{M^{2}_{q}}{M^{2}_{\chi}}, 
\frac{m^{2}_{q^{t},t}}{m^{2}_{\chi^{d},t}},
\frac{M^{2}_{\chi}}{m^{2}_{\rm top}} 
\right) 
+ \frac{m^{2}_{\chi^{d},t}}{m^{2}_{\rm top}}
F_{2}\left( \frac{M^{2}_{q}}{M^{2}_{\chi}}, 
\frac{m^{2}_{q^{t},t}}{m^{2}_{\chi^{d},t}} \right) \right]~,
\label{Tbidoublets}
\eeqa
where $T_{\rm top}$ is the SM contribution due to the top
quark given in Eq.~(\ref{Ttop}), and
\beqa
F_{1}(r_{m}, r_{\lambda}, r_{\rm top}) &=& 2 \left[ \left( r_{m} - 
r_{\lambda} \right) 
\left( 2 \ln r_{\rm top} - 3 \right) - 2 r_{\lambda} \ln r_{m} \right]~,
\\
F_{2}(r_{m}, r_{\lambda}) &=& \frac{4}{3(r_{m} - 1)^{3}} 
\left\{ -3 r_{m} r_{\lambda} (r^{2}_{m} + 1) \ln r_{m} 
\right.
\nonumber \\ & & \left. \mbox
+ (r_{m} - 1) [r^{2}_{\lambda}+ r^{3}_{m} + r^{2}_{m} (r^{2}_{\lambda} + 
3 r_{\lambda} - 2) - r_{m} (2r^{2}_{\lambda} - 3 r_{\lambda} - 1) ] \right\}~.
\eeqa
The functions $F_{1,2}$ are complicated functions of their arguments.
However, the boundary conditions Eqs.~(\ref{Bidoubletsinglet}) imply
that $M_{\chi} < M_{q}$ and $m_{\chi^{d},t} > m_{q^{t},t}$.
Therefore, we concentrate in the region $r_{m} \geq 1$ and $0 \leq
r_{\lambda} \leq 1$.  In this region, both $F_{1}$ and $F_{2}$ are
positive.  Furthermore, given that $m_{\rm top} < M_{\chi}$, one also
has $F_{2} \ll F_{1}$.  In fact, for fixed $0 \leq r_{\lambda} \leq
1$, the ratio $F_{2}/F_{1}$ is bounded by
\beqa
\frac{1-r^{2}_{\lambda}}{6 \ln r_{\rm top} - 9/2} \leq
\frac{F_{2}(r_{m}, r_{\lambda})}{F_{1}(r_{m}, r_{\lambda}, r_{\rm top})} \leq
\frac{1}{6 \ln r_{\rm top} - 9/2}~,
\eeqa
where the lower value is attained for $r_{m} = 1$ and the upper one
for $r_{\lambda} /r_{m} \rightarrow \infty$.  Therefore, in the region
of interest, the second term in Eq.~(\ref{Tbidoublets}) is subdominant
(unless the mixing mass $m_{\chi^{d},t}$ is much larger than $m_{\rm
top}$) and the sign of $T$ is determined by the first term, which is
logarithmically enhanced.  The term associated with $F_{1}$
corresponds to the diagrams shown in
Fig.~\ref{fig:diagramsTbidoublet}.  The logarithm corresponds to an
infrared divergence regulated by the top mass.  Therefore, we have
taken $m_{\rm top} = 167~{\rm GeV}$, the running top mass at the scale
of the pole top mass in our numerical studies.\footnote{A more precise
treatment would integrate out the massive fermion KK modes at the KK
scale of order a few TeV, where the operator $(1/M^{2}_{KK})
H^{\dagger} D_{\mu} H \, \bar{t}^{(0)}_{R} \gamma^{\mu} t^{(0)}_{R}$,
that gives rise to an effective $t_{R}$-$t_{R}$-$W$ vertex, is induced
(as a well as a small 1-loop matching contribution to the
``$T$-parameter'' operator $(1/M^{2}_{KK}) |H^{\dagger} D_{\mu}
H|^{2}$).  The dominant contribution to the $T$ parameter is induced
at the scale of the top mass, where the top quark is integrated out.
$\Delta T$ is the part associated with the effective vertex of $t_{R}$
to the W's above.  Therefore, one should use a running top mass at the
scale where this contribution is induced, but the top Yukawa coupling
that enters the effective $t_{R}$-$t_{R}$-$W$ vertex should be
evaluated at the KK scale.  Our simplified approach errs on the
conservative side.}

For example, in the limit that $M_{\chi} \approx M_{q}$ and
$m_{\chi^{d},t} \approx m_{q^{t},t}$, as is the case when the
zero-modes in the bidoublet are not too close to the IR brane, if we
write
\beqa
M_{\chi} = (1 - \eta_{m}) M_{q}~,
\hspace{1cm}
m_{\chi^d,t} = (1 + \eta_{\lambda}) m_{q^{t},t}~,
\eeqa
and work to first order in $\eta_{m}$ and $\eta_{\lambda}$, we obtain
\beqa
\Delta T = - T_{\rm top} \frac{4 m^2_{\chi^{d},t}}{M^{2}_{q}} 
\left[ 2 (\eta_{m} + \eta_{\lambda}) \ln \frac{M^{2}_{q}}{m^{2}_{\rm top}} -
5 \eta_{m} - 3 \eta_{\lambda} +
{\cal O}\left(\frac{m^{2}_{\chi^{d},t}}{m^{2}_{\rm top}} \, \eta^{2}_{m}, 
\frac{m^{2}_{\chi^{d},t}}{m^{2}_{\rm top}} \, \eta^{2}_{\lambda} 
\right) \right]~.
\label{TBidoubletDegenerate}
\eeqa
We see that in this limit the terms that scale like $m^{4}_{\chi^d,t}$
are suppressed by order $\eta_{m}$ or $\eta_{\lambda}$, and are
negligible unless the mixing $m_{\chi^d,t}$ is sufficiently large
compared to the top mass to overcome the degree of degeneracy
parametrized by the $\eta's$.  The boundary conditions on $q$ and
$\chi$ imply that $\eta_{m}, \eta_{\lambda} > 0$ ($\chi^{d}$ is
lighter and couples more strongly to the Higgs than $q^{t}$).  Thus,
we see that if the logarithm is sufficiently large ($M_{q} \gsim 3.5
m_{\rm top}$), $\Delta T$ is negative.

A different limit is obtained when the zero-modes in the bidoublet are
localized close to the IR brane.  In this case one has $M_{\chi} \ll
M_{q}$ and the expression for $\Delta T$ reduces to
\beqa
\Delta T = - T_{\rm top} \frac{4 m^{2}_{\chi^{d},t}}{M^{2}_{\chi}}
\left[ \ln \frac{M^{2}_{\chi}}{m^{2}_{\rm top}} - \frac{3}{2} - 
\frac{m^{2}_{\chi^{d},t}}{3m^{2}_{\rm top}} 
+ {\cal O}\left( \frac{M^{2}_{\chi}}{M^{2}_{q}} \right) \right]~.
\label{TbidoubletHierarchy}
\eeqa
In fact, in the limit $c_{Q} < -1/2$, $\chi$ becomes ultralight and
its contribution tends to cancel the positive top quark contribution
to the T parameter.  Our analytic expressions assumed $M_{\chi} \gg
m_{\rm top}$ and therefore do not apply in this limit.  However, the
formulas we use in the numerical studies, Eq.~(\ref{ExactT}), do not
suffer from this restriction.  We conclude that in the present class
of scenarios the contribution to the $T$ parameter of the vector-like
bidoublets is always negative.

We turn now to the case where only the first KK mode of the
$SU(2)_{L}$ singlet $t$, with a KK mass $M_{t}$, is retained.  There
is an EW breaking mass that mixes the vector-like singlet with
the zero mode in $q^{t}$, which we call $m_{q^{t}_{0},t}$.  In this
case we obtain the simple result
\beqa
\Delta T = T_{\rm top} \frac{2m^{2}_{q^{t}_{0},t}}{M^{2}_{t}}
\left( \ln \frac{M^{2}_{t}}{m^{2}_{\rm top}} - 1 + 
\frac{m^{2}_{q^{t}_{0},t}}{2m^{2}_{\rm top}} \right)~,
\label{Tsinglet}
\eeqa
which is positive for $M_{t} \gg m_{\rm top}$.  This contribution is
in competition with that of the bidoublets and explains the positive
values of $T$ in the case that the singlet zero-mode is localized away
from the IR brane (smaller values of $c_{t}$ in
Figs.~\ref{fig:TbidoubletsLocal} and \ref{fig:TbidoubletsDelocal}).

We may now explain the qualitative features exhibited in the figures:
\begin{itemize}
\item
The singlet KK mass, $M_{t}$, reaches a minimum for $c_{t} = -1/2$
(the conformal point for a RH zero-mode) and increases
approximately linearly with $c_{t}$ away from that point.  For fixed
$c_{Q}$ the positive contribution due to the singlet,
Eq.~(\ref{Tsinglet}) is maximized near $c_{t} = -1/2$ where $M_{t}$ is
smallest, and is suppressed as $M_{t}$ increases with $c_{t}$.
\item
The rapid increase observed as $c_{t} \rightarrow -1/2$ is due to the
increased value of the 5D top Yukawa coupling, as determined by the
top mass, $m_{\rm top}$.  This enhances the effects due to mixing via
the Higgs vev [e.g., the second term in Eq.~(\ref{Tbidoublets})].
\item 
The KK bidoublets become more degenerate as $c_{Q}$ moves in the
positive direction, so their negative contribution to T becomes
smaller [see Eq.~(\ref{TBidoubletDegenerate})].  As $c_{Q}$ approaches
$-1/2$, the negative contribution due to the light mode $\chi^{d}$
becomes more important (solid curve in Fig.~\ref{fig:TbidoubletsLocal}).
\end{itemize}
The quantitative behavior also depends on the EW breaking masses that
mix the first KK modes of the bidoublets and the singlet, that were
not included in our analytic expressions above.  However,
Fig.~\ref{fig:TbidoubletsLocal} includes these effects exactly, as
well as the effects due to higher KK modes (which are negligibly
small).

\begin{figure}[t]  
\centerline{\includegraphics[width=0.8\textwidth]{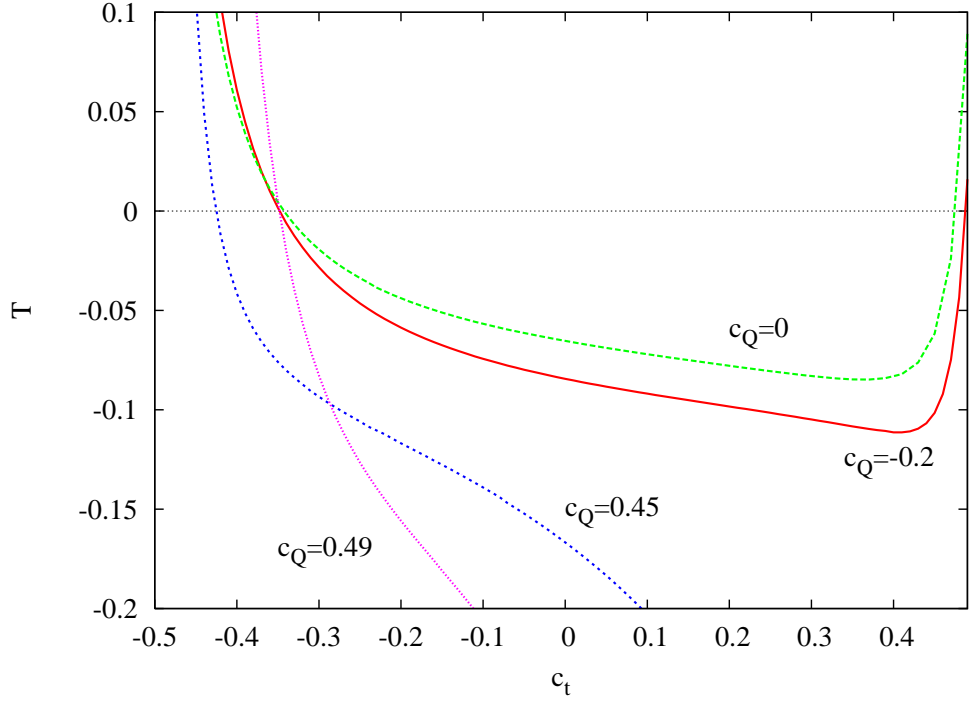}} 
\caption{Contribution to the $T$ parameter involving the KK modes of
the bidoublets and triplets which couple to the Higgs through the
top Yukawa coupling, for the parity and quantum number assignments 
given in    Eq.~(\ref{Bidoublettriplet}), 
We use $\tilde{k} = 1.5~{\rm TeV}$ and  
$m_{\rm top} = 167~{\rm GeV}$.  It is
assumed that the Higgs has the profile of gauge-Higgs unification
models, Eq.~(\ref{higgs:norm}).}
\label{fig:TDecuplet}  
\end{figure}  
For the cases that include triplets, as in
Eq.~(\ref{Bidoublettriplet}), Eq.~(\ref{ExactT}) has to be generalized
to include the mixing within states of charge $5/3$, $2/3$ or $-1/3$.
We show the $T$ parameter for this case in Fig.~\ref{fig:TDecuplet}.
The new ingredient is the presence of states, not part of a bidoublet,
with parity assignments $(+,-)$ or $(-,+)$.  This may lead to additional
light states~\cite{Agashe:2004bm}, 
and correspondingly important contributions to the $T$
parameter, depending on the value of $c_{t}$
(the discrete parity $P_{LR}$
implies that both triplets, $T_{1}$ and $T_{2}$, are controlled by the
same localization parameter, which we call again $c_{t}$).  For the
parity assignments of the triplets in Eq.~(\ref{Bidoublettriplet}),
that were motivated by the $SU(2)_{R} \times P_{LR}$ symmetry
protecting $T$ and $\delta g_{b\,L}/g_{b\,L}$, the triplet states
become light as $c_{t} \rightarrow 1/2$ (the RH zero mode is
localized near the IR brane, and the first KK mode obeying $(-,+)$
boundary conditions for its RH components becomes light).
These give a positive contribution to the $T$ parameter that explains
the upward turn in the curves as $c_{t} \rightarrow 1/2$.  In
particular, the $T$ parameter can be positive in this region.  Away
from this region the triplet states are not particularly light and the
behavior is similar to what was found for the assignments of
Eq.~(\ref{Bidoubletsinglet}): $T$ is dominated by a negative
contribution due to the bidoublets, except near $c_{t} \sim -1/2$ (the
conformal point for RH modes) where a large 5D Yukawa
coupling enhances the mixing effects and the importance of various
positive contributions (see the second term in
Eq.~(\ref{Tbidoublets}), for example).

Thus, we see that the mechanism suggested in \cite{Agashe:2006at} to
control the anomalous couplings of the bottom to the $Z$ implies a
sizable negative contribution to the $T$ parameter in large regions of
parameter space.  There are, however, regions where $T$ can be
positive.  Given the positive $S$ parameter discussed in section
\ref{sec:gaugecorrections}, and that under these circumstances the EW
data prefer positive values of the $T$ parameter, we conclude that the
favored region has $c_{Q}$ not too close to $1/2$ and $c_{t}$
somewhere around $-0.4$ to $-0.5$ 
[or also around $0.5$ for models based on the
assignment (\ref{Bidoublettriplet})].~\footnote{Note that a negative
  value of the $S$ parameter as might be obtained with the mechanism
  of Ref.~\cite{Hirn:2006nt} could be compatible with a negative value of the $T$
parameter.}  We postpone a more detailed
discussion of the EW constraints to the next section, where very
similar bounds are found in the context of gauge-Higgs unification
models.  We will see that gauge bosons with KK masses around $3~{\rm
TeV}$ are allowed.  The most important lesson of this section is that
the contribution to $T$ can be sufficiently important to select rather
well-defined regions of parameter space, in this case the localization
of the top quark multiplets in models based on $SU(2)_{L} \times
SU(2)_{R} \times P_{LR}.$ 

Let us finish this section by mentioning
that vector-like quarks as the ones present in these models also
contribute to the $S$ parameter at one
loop. 
Such a contribution is always positive and
much less dependent on the parameters of the model. It is given by
\beqa
\label{Sloop}
S &=& \frac{N_c}{4\pi}
\sum_{\alpha,\beta} \left[
\left(
U^{\dagger \alpha\beta}_L Y_L^{\beta\alpha} +
U^{\dagger \alpha\beta}_R Y_R^{\beta\alpha} \right)
\bar{\chi}_{+}(\MM^\mathrm{all}_{\alpha\alpha},\MM^\mathrm{all}_{\beta\beta})
\right.
\nonumber \\
& & \hspace{3.7cm} \left. \mbox{} +
\left(
U^{\dagger \alpha\beta}_L Y_R^{\beta\alpha} +
U^{\dagger \alpha\beta}_R Y_L^{\beta\alpha} \right)
\bar{\chi}_{-}(\MM^\mathrm{all}_{\alpha\alpha},\MM^\mathrm{all}_{\beta\beta}) 
\right]~,
\eeqa
where
\beqa
\bar{\chi}_+(y_1,y_2) &=& \frac{5(y_1^4 + y_2^4) - 22 y_1^2 y_2^2}
{9 (y_1^2 - y_2^2)^2} +
\frac{3 y_1^2 y_2^2 (y_1^2 + y_2^2) - (y_1^6 + y_2^6)}{3 (y_1^2 -
y_2^2)^3}
\ln\left(\frac{y_1^2}{y_2^2}\right)
-\frac{2}{3} \ln \left(\frac{y_1 y_2}{\mu^2}\right)~,
\nonumber \\ [.7 em]
\bar{\chi}_{-}(y_1,y_2) &=&
\frac{y_1 y_2}{(y_1^2 - y_2^2)^3} \left[ y_1^4 - y_2^4 -
2 y_1^2 y_2^2 \ln \left(\frac{y_1^2}{y_2^2} \right) \right]~.
\eeqa
We assume that there are no exactly massless fermions in the theory.
These would lead to IR divergences that need to be carefully
subtracted. In the above, $\mu$ is an arbitrary scale that cancels out
as a consequence of ${\rm Tr} [U^{\dagger}_{L} Y_{L} + U^{\dagger}_{R}
Y_{R}] = 0$,
where $Y_{L,R}$ is the matrix of left- and right-handed
hypercharge couplings in the mass eigenstate basis, while $U_{L,R}$ is
the corresponding matrix of couplings to $W^{3}_{\mu}$. Note that in
Eq.~(\ref{Sloop}) we have put \textit{all} 
fermion masses in a single (diagonal) matrix $\MM^\mathrm{all}$.
To obtain the contribution due to the new physics one needs to subtract
the SM part. For example, the one due to the top-bottom system is
\beqa
S_{\rm t,b} = - \frac{1}{6\pi} \left[ \ln\left(\frac{m_{\rm
top}^2}{m_{\rm bottom}^2} \right) - 3 \right]~.
\eeqa
We will include the
phenomenological impact of this contribution to $S$ in the analysis of
the next section.

%-----------------------------------------------------------------------------
\section{Gauge-Higgs Unification}
\label{sec:gaugHiggs}
%-----------------------------------------------------------------------------

The models based on $SU(2)_{L} \times SU(2)_{R} \times U(1)_{X}$
discussed in the previous sections can be naturally embedded into
$SO(5)\times U(1)_{X}$.  The analysis of
subsection~\ref{sec:gaugecorrections} applies to this case by simply
taking $g_{L} = g_{R}$.  The additional gauge fields have the quantum
numbers of the SM Higgs under $SU(2)_{L} \times SU(2)_{R}$ and offer
the interesting possibility that the Higgs be part of a higher
dimensional gauge field.  This can help address the little hierarchy
problem present in RS scenarios based on a fundamental scalar Higgs
field.  Imposing $(+,+)$ boundary conditions on the fifth component of
the gauge fields in $SO(5)/SU(2)_{L} \times SU(2)_{R}$ allows one to
identify the SM Higgs with their zero-mode components.  One finds that
these zero-modes have a nontrivial profile along the extra dimension
\beqa
A^{\hat{a}}_5(x,y) &=& A^{\hat{a} (0)}_5(x) f_H(y) + \cdots 
\label{higgs:A5}
\eeqa
where $\hat{a}$ labels the generators of $SO(5)/SU(2)_{L} \times
SU(2)_{R}$, and
\beq
f_H(y) = \left(\frac{2k}{e^{2kL}-1}\right)^{1/2} \, e^{2ky}~.
\label{higgs:norm}
\eeq
The Higgs field corresponds to the $SU(2)_{L} \times SU(2)_{R}$
bidoublet contained in the adjoint representation of $SO(5)$.  

Regarding the fermion sector, the bidoublet plus singlet system of
Eq.~(\ref{Bidoubletsinglet}) fits into the fundamental representation
of $SO(5)$,
\beqa
5  &\sim& (2,2) \oplus 1~,
\eeqa
while the system of a bidoublet and triplets of
Eq.~(\ref{Bidoublettriplet}) fits into the 10-dimensional
representation of $SO(5)$,
\beqa
10 &\sim& (2,2) \oplus (3,1) \oplus (1,3)~.
\eeqa
Therefore, no fermions with quantum numbers different to those already
encountered in the $SU(2)_{L} \times SU(2)_{R}$ theory studied in the
previous sections are necessary.  The results for the fermion loop
contributions to the $T$ parameter can then be understood from the
physics already discussed, with a couple of restrictions:
\begin{itemize}
\item
A single bulk mass parameter controls simultaneously the localization
of the bidoublet and singlet (or triplets) in a given $SO(5)$
multiplet.
\item
The Yukawa couplings are no longer free parameters, but are related to
observed gauge couplings.
\end{itemize}

The SM fermions can be embedded in various ways into the above
5-dimensional $SO(5)$ structure.  An economical way to do it is to let
the SM singlet and doublet components of, say, the up-type sector arise
from a single $SO(5)$ multiplet.  That is, let the multiplets in
Eqs.~(\ref{Bidoubletsinglet}) or (\ref{Bidoublettriplet}) come from
the same $SO(5)$ multiplet:
\beqa
\begin{array}{rclrclrcl}
\xi^{5} &=& 
Q & \oplus & u
\end{array}~,
\eeqa 
or
\beqa
\begin{array}{rclrclrcl}
\xi^{10} &=& 
Q & \oplus & T_{1} & \oplus & T_{2} 
\end{array}~.
\eeqa 
In either case, the down-type sector must be assigned to a different
$10$ of $SO(5)$.  In these scenarios the up-type Yukawa couplings
arise directly from the bulk gauge interactions, while the down-type
Yukawa couplings can arise from mixing effects such as those discussed
below.

It is instructive to consider the expression for the top Yukawa
coupling in these cases.  When both the LH and RH
components of the top quark come from a $5$ of $SO(5)$, one finds
\beq
\lambda^{5}_{\rm top} = \frac{g_{4}}{\sqrt{2}} 
\left[ \frac{(\frac{1}{4} - c^{2})2kL e^{2kL}}{(1 - e^{(1-2c)kL}) 
(1 - e^{(1+2c)kL})} 
\right]^{1/2}~,
\label{TopYukawaSimple}
\eeq
where the factor of $1/\sqrt{2}$ arises from the normalization of the
Higgs field.  The factor in parenthesis arises from the integral over
the extra dimension of the Higgs wavefunction, Eq.~(\ref{higgs:norm}),
times the fermion zero-mode wavefunctions ($c$ is the dimensionless
mass parameter that controls the localization of the zero-modes).
This factor reaches a maximum at $c=0$, and is exponentially
suppressed for $|c| > 1/2$.  Taking $g_{4} \approx 0.65$, a 
running top quark mass
of $167~{\rm GeV}$ is obtained for $|c| \approx 0.42$.  For $c =
0.42$, we can directly read the associated contribution to the $T$
parameter from Fig.~\ref{fig:TbidoubletsDelocal} by setting $c_{1} =
c_{2} = 0.42$.  For $\tilde{k} = 1.5~{\rm TeV}$, which corresponds to
gauge KK masses starting at $3.75~{\rm TeV}$, this gives $\Delta T
\approx -0.08$.  For $c = -0.42$ one finds $\Delta T \approx -0.25$.

If the top quark comes from a decouplet, one finds instead
\beqa
\lambda^{10}_{\rm top} = \frac{1}{2\sqrt{2}} \lambda^{5}_{\rm top}~.
\eeqa
The factor of $1/2$ is an $SO(5)$ group theory factor, while the
$1/\sqrt{2}$ comes from the normalization of the isospin-$0$ component
of the $SU(2)_{R}$ triplet.  For $c = 0$, the maximum obtainable top
mass is about $110~{\rm GeV}$, so that this scenario is ruled out.

A different possibility is to assign an independent $SO(5)$ multiplet
for each SM model fermion.  In this case, the gauge coupling, which is
diagonal in flavor space, does not induce masses for the zero-modes.
However, localized mass terms that mix different $SO(5)$ multiplets
can generate the fermion masses and mixing angles, once the Higgs
field
gets a vev.  Since 
the Higgs field is
localized near the IR brane, see  Eq.~(\ref{higgs:norm}), 
only masses localized on the IR brane can
be effective, and we may restrict ourselves to such cases.  
These masses preserve
$SU(2)_{L} \times SU(2)_{R}$, so one may have localized masses 
that we call $\MQ$ for the bidoublets, $\Mu$ for the singlets, and 
$\hat{M}_{T_{1}}$ and $\hat{M}_{T_{2}}$ for the triplets.  Due to the
localization, these ``mass parameters'' are dimensionless.

We are mainly interested in the fermion loop contributions to
the $T$ parameter, thus we restrict ourselves 
again to the top quark sector (the
contributions due to the other quark and lepton fields can be made
negligibly small).  A priori there are several ways of assigning the
LH and RH top quark components to the $5$ and $10$
representations of $SO(5)$.  However, due to the group theory factors
associated with the 10-dimensional representation of $SO(5)$ it is
impossible to accommodate the observed top mass unless the $SU(2)_{L}$
singlet top comes from a $5$.  The $SU(2)_{L}$ doublet components can
arise either from a $5$ or a $10$ of $SO(5)$.\footnote{When both a $5$
and a $10$ are used, only the bidoublets in each multiplet can mix,
due to the unbroken $SU(2)_{L} \times SU(2)_{R}$ symmetry on the IR
brane.  Maximizing the top Yukawa coupling requires maximizing the
mixing.  If one assigns the singlet to the $10$ and makes the mixing
large, one finds a situation where effectively both chiralities come
from a $10$, which leads to a small top mass.  Therefore, the singlet
must come from a $5$.} Therefore, in the top quark sector, only the
bidoublet and singlet localized masses ($\MQ$ and $\Mu$) are
relevant.

The localized masses also affect the spectrum of KK modes and have the
important consequence that they make the light states even lighter.
Based on this observation we see that
\begin{itemize}
\item
The bidoublet localized mass, $\MQ$, by pushing the $\chi$ states to
lower masses, has the effect of enhancing the negative contributions
to the $T$ parameter discussed in section \ref{sec:Tparameter}.
\item
The localized mass $\Mu$ can also generate light states in the
singlet towers, which in general enhances their positive contributions
to the $T$ parameter.
\end{itemize}
We find that whenever the bidoublet mass, $\MQ$, is appreciable, the
negative contributions to the $T$ parameter are very important.  In
fact, if the top mass is generated only from $\MQ$, $T$ is negative
for all $c_{1}$ and $c_{2}$, the localization parameters for the two
$SO(5)$ multiplets that generate the top quark \cite{gaugeHiggs}.

Given the restrictions imposed by the top quark mass, and in order
to obtain positive values of $T$, we 
consider the case with only quintuplets of $SO(5)$,
and choose the parities so that bidoublet mixing masses
are forbidden:
\beqa
\begin{array}{rcl}
\xi_{1L} &\sim& 
\begin{array}{rclrcl}
Q_{1L} &=& \begin{pmatrix} 
\chi^{u}_{1L}(-,+) & q^{u}_{L}(+,+) \\
\chi^{d}_{1L}(-,+) & q^{d}_{L}(+,+)
\end{pmatrix}
& \oplus &
u^{\prime}_{L} (-,+)
\end{array}~,
\\ [1.5em]
\xi_{2R} &\sim& 
\begin{array}{rclrcl}
Q_{2R} &=& \begin{pmatrix} 
\chi^{u}_{2R}(+,-) & q^{\prime u}_{R}(+,-) \\
\chi^{d}_{2R}(+,-) & q^{\prime d}_{R}(+,-)
\end{pmatrix}
& \oplus &
u_{R} (+,+)
\end{array}~,
\end{array}
\label{ParityqQ}
\eeqa
where, under $SU(2)_{L} \times SU(2)_{R}$, $Q_{i} \sim (2,2)$ for
$i=1,2$, and $u$ and $u'$ are singlets under this symmetry.  All multiplets
are taken to have charge $X = 2/3$.  The parities of $Q_{1}$ and $u$
are fixed by the unbroken $SU(2)_{R}$ symmetry on the IR brane, and by
the low-energy content.  The parities of $Q_{2}$ are chosen so the
bidoublets cannot mix through IR brane localized masses, and the
parity of $u^{\prime}$ is then fixed so a singlet mixing mass, that
generates the top quark mass, can be written:
\beq
\delta(L-y) \left[ \Mu \bar{u}^{\prime}_{L} u_{R} 
+ {\rm h.c.} \right]~.
\label{localizedmasses}
\eeq
The parities of the remaining chiralities are opposite to the ones
shown in Eq.~(\ref{ParityqQ}).

This system is relatively simple to analyze.  There are three
fundamental parameters, $c_{1}$, $c_{2}$ and $\Mu$.  Given $c_{1}$ and
$c_{2}$, the top mass fixes $\Mu$. The top Yukawa coupling is given
by 
\beqa
\lambda_{\rm top} &\approx& \frac{g_{4}}{\sqrt{2}} \Mu 
\left[\frac{(\frac{1}{2} - c_{1})(\frac{1}{2} + c_{2})2kL 
e^{2(1+c_2-c_1)kL}}{(1 
- e^{(1-2c_{1})kL}) 
(1 - e^{(1+2c_{2})kL})} \right]^{1/2}
\nonumber \\
&& \quad \quad \quad \quad \mbox{}\times 
\left[1 + 
\Mu^2 \; e^{2(c_2-c_1)kL} \;
\frac{(\frac{1}{2}+c_2)(1-e^{(1+2 c_1)kL})}{(\frac{1}{2}+c_1)
(1-e^{(1+2 c_2)kL})} 
\right]^{-1/2},
\eeqa
In the limit of small $\Mu$,
the top Yukawa coupling is given by
\beq
\lambda_{\rm top} \approx \frac{g_{4}}{\sqrt{2}} \Mu 
\left[ \frac{(\frac{1}{2} - c_{1})(\frac{1}{2} + c_{2})2kL 
e^{2(1+c_2-c_1)kL}}{(1 
- e^{(1-2c_{1})kL}) 
(1 - e^{(1+2c_{2})kL})} 
\right]^{1/2}~,
\eeq
while for $\Mu \gg 1$ the singlet zero-mode lives in $\xi_{1}$, and
the top Yukawa coupling reduces to Eq.~(\ref{TopYukawaSimple}) with $c =
c_{1}$.  In particular, it becomes independent of $c_{2}$.  For given
$c_{1,2}$, this last limit corresponds to the maximum achievable top
Yukawa coupling.  If $|c_{1}|$ is too large, it may be impossible to
accommodate a large enough Yukawa coupling (for example, for $m_{\rm
top} = 167~{\rm GeV}$, one needs $|c_{1}| \lsim 0.42$, although mixing
with light KK states can change this.)  Therefore, the system contains
two free parameters $c_{1}$ and $c_{2}$, where $|c_{1}|$ cannot be too
close to $1/2$ in order to generate a sufficiently large top Yukawa
coupling.

\begin{figure}[t]  
\centerline{\includegraphics[width=0.8\textwidth]{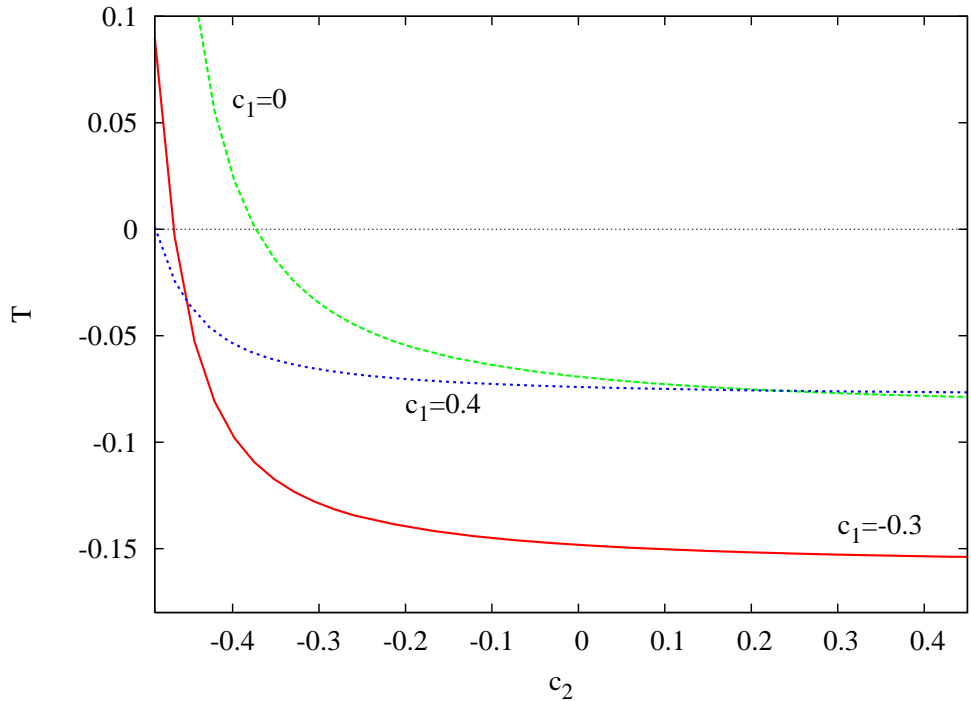}}
\caption{Contribution to the $T$ parameter involving the KK modes of
Eq.~(\ref{ParityqQ}), which couple to the Higgs through the
top Yukawa coupling.  We use $\tilde{k} = 1.5~{\rm TeV}$ and $m_{\rm
top} = 167~{\rm GeV}$.}
\label{fig:TSO5SImple}  
\end{figure}  
In Fig.~\ref{fig:TSO5SImple} we plot the $T$ parameter as a function
of $c_{2}$ for several values of $c_{1}$.  We see that the situation
is qualitatively similar to the case without gauge-Higgs unification
presented in Fig~\ref{fig:TbidoubletsDelocal}, with $T$ negative for
most values of $c_{2}$, and a rapid increase as $c_{2}$ approaches
$-1/2$.\footnote{As we have mentioned, the bottom quark must arise
from a different multiplet, $\xi_{3}$, that is a $10$ of $SO(5)$.
Therefore, the bottom quark mass is generated through a localized
mixing mass, $\MQ$, for the bidoublets in $\xi_{1}$ and $\xi_{3}$.
This mass makes the $\chi^{d}$'s in $\xi_{1}$ lighter and enhances
their negative contribution to $T$.  Therefore, it should be taken
sufficiently small for a region with positive (or zero) $T$ to exist
\cite{gaugeHiggs}.}
\begin{figure}[t]  
\centerline{\includegraphics[width=0.8\textwidth]{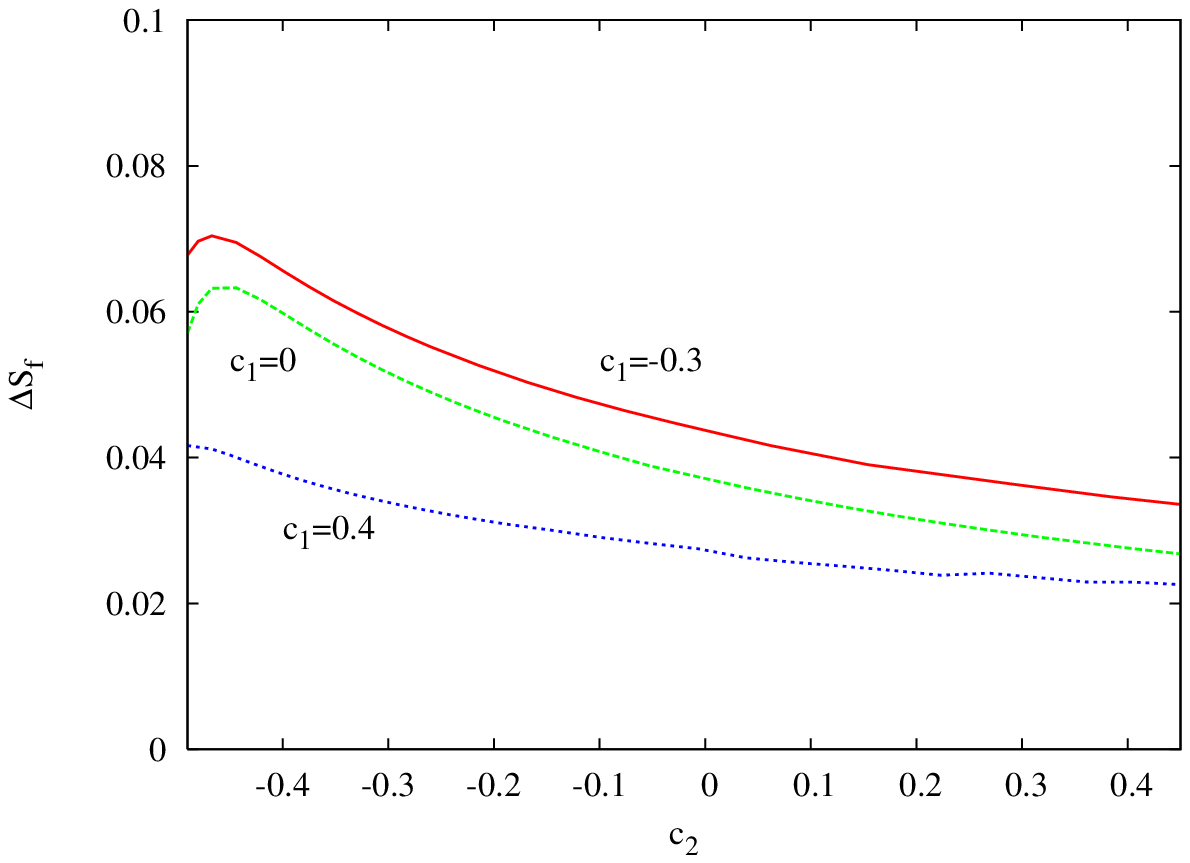}} 
\caption{Contribution to the $S$ parameter involving the KK modes of
Eq.~(\ref{ParityqQ}), which couple to the Higgs through the
top Yukawa coupling.  We use $\tilde{k} = 1.2~{\rm TeV}$ and $m_{\rm
top} = 167~{\rm GeV}$.}
\label{fig:SSO5SImple}  
\end{figure}  

Given a positive $S$ parameter, and for small values of $U$, as is 
the case when the first two
families are localized near the UV brane, 
the EW data prefer a positive value of $T$.  
We see in Fig.~\ref{fig:TSO5SImple} 
that this restricts $c_{2}$ to be in the
vicinity of $-0.4$ to $-0.5$, where $T$ crosses through zero.  Thus,
the most important constraints in the previous scenario come from the $T$ 
and 
$S$ parameters discussed in section~\ref{sec:gaugecorrections}, plus
the extra contribution from the loops of fermion KK modes, Eq.~(\ref{Sloop}).  
With
the first two families near the UV brane (localization parameters for
LH zero-modes, $c \gsim 0.55$ or so), the prediction for $S$ is
\beqa
S \approx 9 \, \frac{v^{2}}{\kk^{2}} 
+ \Delta S_f~,
\eeqa
where $\kk = k e^{-kL}$, and $\Delta S_f$
is the contribution from the fermion loops given in
Eq.~(\ref{Sloop}). In Fig.~\ref{fig:SSO5SImple} we show 
$\Delta S_f$ as a function of $c_{2}$ for several values of $c_{1}$.
We see that, as we said, it is positive and much less dependent on
the parameters of the model.

For a light Higgs 
with $m_{H} \simeq 115~{\rm GeV}$ (recall that gauge-Higgs unification
models typically predict a light Higgs), a $2\sigma$
bound on $S \lsim 0.3$
appears~\cite{unknown:2005em}. 
In order to be consistent
with the $2\sigma$ $S$-$T$ bounds for the largest allowed values of $S$, 
a positive contribution of $T
\approx 0.3$ is also required.  
The bound on $S$ leads to 
a lower bound $\kk \approx 1.2~{\rm TeV}$ (this includes a
contribution $\Delta S_f \approx +0.06$), which corresponds to
KK gauge boson masses of $M_{KK}\approx 2.5 \, \kk \approx 3~{\rm
TeV}$.  In turn, the positive contribution to $T$ can arise from
the 1-loop effects associated with the top sector discussed above,
for specific values of the bulk mass parameters.
For example, taking $c_{1} = 0$, this can be obtained for $c_{2} =
-0.468$.  The top mass fixes $\Mu \approx 2.91$.  The gauge contributions to the $T$ and $U$ parameters of
Eq.~(\ref{STU}) are negligible ($\Delta T_{\rm gauge} \approx -0.006$
and $\Delta U_{\rm gauge} \approx 0.005$). 

For the above values of parameters, one also finds $\delta
g_{b\,L}/g_{b\,L} \approx 0.8 \times 10^{-3}$, arising from the gauge
contribution in  
Eq.~(\ref{epsilonb:bidobletes}). There are also potentially important
loop-level contributions to 
$\delta g_{b\,L}/g_{b\,L}$, involving the sector of fermion  KK modes
that mix with the top. In fact, in order to obtain a positive $T$
parameter this mixing needs to be relatively large. We have estimated
this contribution using the results of \cite{Bamert:1996px}, and
obtained $\delta g^{\rm loop}_{b\,L}/g_{b\,L} \approx -5 \times
10^{-3}$.  The net value of $\delta g_{b\,L}/g_{b\,L}$ falls outside
the range given in Eq.~(\ref{epsilonb-bound}).
A naive estimation, using only $R_b$, would seem to push the lower
bound 
$\tilde{k}\approx
1.5$ TeV. A more careful analysis requires a global fit to all EW
observables that is currently underway~\cite{gaugeHiggs}.

%, which assumes $\delta
%g_{b\,R} = 0$. However, as noted in Ref.~\cite{Haber:1999zh,Beauty} a
%non-vanishing value  of $\delta g_{b\,R}$ modifies this conclusion.  
%Depending on the localization of the right-handed bottom, $\delta
%g_{b\,R}$ can receive sizable contributions from the gauge sector,
%analogous to Eq.~(\ref{epsilonb}). As mentioned before, the
%right-handed bottom arises from an $SO(5)$ decouplet with $U(1)_{X}$
%charge $2/3$, and transforms as a triplet under
%$SU(2)_{R}$. Considering $T^3_R = -1$ and  $T^3_L = 0$ for the
%$b_{R}$, we find 
%%
%\begin{equation}
%\frac{\delta g_{b\,R}}{g_{b\,R}} 
%\approx \frac{e^2}{s^2 c^2} 
%\left[
%G^{b_{R}}_{++} + 
%9 \, G^{b_{R}}_{-+} - 
%G^f_{++} \right] ~.
%\label{epsilonbR:bidoublets}
%\end{equation} 
%%
%For example, taking $c_{b_{R}} \approx -0.42$ yields $\delta g_{b\,R}
%\approx -0.17$, which leads to an improvement in the agreement with
%the measured values of $R_{b}$ and the bottom forward-backward
%asymmetry.\footnote{Notice that tree-level mixing effects of the
%  bottom with charge $-1/3$ states are small, either because the
%  states are heavy or because their mixing is suppressed by the bottom
%  mass.} 

%For our choice of field multiplets,
%Eq.~(\ref{ParityqQ}), there are no vector-like quarks that can mix
%with the LH bottom.  
We conclude that there are rather well
defined regions of parameter space that pass all current bounds from
precision measurements with relatively light, and probably accessible
at the LHC, gauge boson KK excitations.  In fact, one typically finds
fermion KK excitations, with masses below 1~TeV, 
that should give clear signals at the LHC, as we discuss next.

%-----------------------------------------------------------------------------
\section{Phenomenology}
\label{phenomenology}
%-----------------------------------------------------------------------------

We have seen in the previous sections how the $SU(2)$ custodial
symmetry plus the $L\leftrightarrow R$ discrete symmetry allow for
Randall-Sundrum models with KK excitations of the gauge bosons as low
as $M_{KK} \sim 3$ TeV, mainly constrained by the $S$ parameter. 
In order to avoid large negative contributions to the $T$ parameter at
the loop level, a very well defined pattern emerges, with the RH top
quark not so strongly localized near the IR brane ($c_{2} \sim
-0.4$ to $-0.5$, $-0.5$ being the conformal point for a RH fermion). 
We will focus on the gauge-Higgs unification model discussed in the
previous section, with the quantum
numbers and parities as in Eq.~(\ref{ParityqQ}), and will outline the
differences for other models at the end.
The net positive contribution to the $T$ parameter in this case
comes from a large positive contribution from a vector-like
$SU(2)_L$ singlet that compensates a relatively large negative
contribution from a  light bidoublet
($c_{2} \sim -1/2$ localizes a LH zero mode near the IR
brane, and the LH components of $Q_{2}$, that obey $(-,+)$
boundary conditions develop a light mode). Furthermore, the singlet is
lighter than one would have expected from its localization
parameter $c_2$ due to the effect of the localized mass $\hat{M}_u$.
As an illustration, for the choice of parameters satisfying the EW
constraints discussed above, one finds for the states with charges
$5/3$ and $-1/3$, that do not mix through the Higgs vev: 
\beqa
M_{\chi^{u}_2} = M_{q^{\prime d}} &\approx& 470~{\rm GeV}~.
\eeqa
The lightest charge $2/3$ states, coming from $Q_{2}$, are split due
to EW symmetry breaking effects. In fact, only one linear combination of
$\chi^{d}_{2}$ and $q^{\prime u}$ mixes with the singlets. We call
$q_{1}$ the linear combination that does not feel
the Higgs vev. We call the two remaining states $q_{2}$ and $u_{2}$
(for a small Higgs vev, $q_{2}$ would be mostly an $SU(2)_{L}$ doublet
while $u_{2}$ would be mostly an $SU(2)_{L}$ singlet). Their masses are  
\beqa
M_{q_{1}} \approx 470~{\rm GeV}~,
\hspace{1cm}
M_{q_{2}} \approx 495~{\rm GeV}~,
\hspace{1cm}
M_{u_{2}} \approx 742~{\rm GeV}~.
\eeqa
The mixing between the top quark and the latter mode is about $\sim 32\%$.
There are additional fermion states starting at about $1.9~{\rm TeV}$.

Light vector-like quarks that mix strongly with the top have two main
phenomenological effects. First, anomalous couplings of the SM
quarks to the $Z$ and $W$ bosons~\cite{delAguila:2000rc} are induced
due to mixing. For instance, in the numerical example we are
considering the LH top coupling to the $Z$ and the $Wt_L b_L$ coupling
are reduced by 
\begin{equation}
\delta g_{Ztt}^L/g_{Ztt}^L \sim -0.2~,
\quad \delta g_{Wtb}^L/g_{Wtb}^L \sim -0.07~,
\end{equation}
The $W t_L b_L$  coupling has not been directly measured 
yet. Early studies for single top production at the 
Tevatron~\cite{Amidei:1996dt} show that it may be measured with
a precision of about 10~\% for a total integrated luminosity of
order 8~fb$^{-1}$. At the LHC, this coupling
is expected to be measured to a $\sim 5$--$10\%$ 
precision~\cite{Altarelli:2000ye}. 
Apart from the effect of the mixing with the fermion KK modes, there
is an extra effect due to the mixing of the $Z$ with the
\textit{gauge} KK modes, similar to the one we discussed for the
bottom couplings in section~\ref{sec:gaugecorrections}. Actually, the
mechanism that protects the bottom quark coupling from corrections
cannot simultaneously protect the top couplings~\cite{Agashe:2006at}. 
Eq.~(\ref{epsilonb}) applies equally well to the top quark where we
now have $T^3_R=-T^3_L=1(0)$ for the LH (RH) top quark
chirality. This results in a very small modification of the RH
coupling (recall that the one loop contribution to the $T$ parameter
requires the RH top to leave not so close to the IR brane) whereas the
LH coupling is modified by a factor
\begin{equation}
\delta g^{L (\mathrm{gauge})}_{Ztt}/g^L_{Ztt} \sim -0.04~,
\end{equation}
whereas the $W t_L b_L$ coupling receives a correction
\begin{equation}
\delta g^{L (\mathrm{gauge})}_{Wtb}/g^L_{Wtb} \sim -0.015~.
\end{equation}
As we see, this effect is smaller than the one coming from mixing with
fermion KK modes. Apart from modifications of the diagonal top
couplings, Flavor Changing Neutral Couplings (FCNC) are also generated
both through mixing with vector-like quarks and due to the different
top quark couplings in the gauge eigenstate basis that get mixed in
the physical basis. These effects depend on the mixing with the first
two families that we have not considered in detail here. The latter
effect has been recently discussed in~\cite{Agashe:2006wa} with the
result that FCNC possibly observable at the LHC can be generated. 
Recall, however, that due to the constraints associated with the $T$
parameter it is the LH top that is localized closer to the IR brane, and
therefore it receives the largest modifications to its couplings. The
authors of
Ref.~\cite{Agashe:2006wa}, instead, assumed that the largest effects 
appear in the RH sector.

Another phenomenological implication of light vector-like fermions is
of course the possibility of direct production at colliders.
The two relevant modes that
are light and mix with the top are, in the gauge eigenstate basis, 
a bidoublet that is an equal
admixture of a quark with isospin $+1/2$ and isospin $-1/2$ (and
therefore its effects due to mixing are typically suppressed) and a
singlet that mixes strongly with the top. Thus, as a first
approximation we will consider the phenomenology of just the singlet. A
more detailed analysis of the collider phenomenology of these and 
other fermion KK modes will be presented elsewhere~\cite{gaugeHiggs}. 

Vector-like singlets of charge $2/3$ can be pair or singly produced at
colliders. Pair production is quite independent of the heavy quark mixing
with the top but the cross section dies off very quickly with the mass
of the heavy quark. For instance, the production cross section of a
500~GeV quark at the Tevatron collider is about 1~fb~\cite{Rosner},
while it increases to values of a few~pb at the LHC. Such small
cross sections at the Tevatron imply that searches for these quarks,
decaying mainly into third generation states, become quite 
challenging~\cite{Rosner,Morrissey:2003sc}.
Searches become much more promising at the LHC.
A recent
study~\cite{Aguilar-Saavedra:2005pv} shows that a 500 GeV (1 TeV) 
vector-like quark singlet can be discovered at the LHC with an integrated
luminosity of $\sim 1.2$ fb$^{-1}$ ($\sim 90$ fb$^{-1}$) thus rendering
the signatures of our model easily observable.

Single production is not suppressed as much by the
mass of the heavy quark but depends on the details of the mixing with
the top-bottom sector.
A comparison with the ATLAS study of little Higgs
models~\cite{T:littleHiggs} shows that a vector-like singlet with
the parameters of our numerical example is well within LHC reach in
single production. In particular the $T \to W b$ channel looks
particularly promising in the discovery of a quark with mass of order
$750$ GeV and a
$\sim 32\%$ mixing with the top.
Furthermore, the mass of the singlet is typically light enough to
make pair production competitive with single production. 

We have thus seen that realistic models of gauge-Higgs unification
with gauge KK excitations with masses as low as $3$ TeV 
can be constructed. Generic predictions of these
models include deviations of the $Wtb$ coupling sufficiently large to be 
observed  with the $\sim 5$--$10\%$ accuracy achievable at the
 LHC, as well as vector-like quarks light enough to be 
  produced at the LHC, both in pairs and together with SM quarks.
 Alternative choices of
quantum numbers and parities, but still compatible with EW
precision observables, can have slightly different features regarding
the spectrum of bidoublets, but light and strongly mixed vector-like
singlets remain as a solid prediction of the models. If we relax the
gauge-Higgs unification condition and allow for a fundamental Higgs as
in sections~\ref{sec:LRmodels} and~\ref{sec:Tparameter},
then vector-like singlets do not need to be that light.
However, they
still need to mix very strongly with the top to compensate for the
negative contribution of bidoublets to the $T$ parameter. This
typically results
in a scenario with somewhat light
vector-like bidoublets, with masses $M_\chi
\sim 1$ TeV, and heavy vector-like singlets, with masses $M_t\sim
2.5$ TeV, which will be more challenging for the LHC.
However, due to the strong mixing, 
large corrections to the top gauge couplings,
in the $\sim 10-20\%$ range, and therefore observable at the LHC, 
are expected in these models.

%----------------------------------------------------------------------------
\section{Conclusions}  
\label{sec:conclusion}  
%----------------------------------------------------------------------------  

The Randall-Sundrum scenario with gauge and fermion fields propagating
in the bulk offers an attractive solution to the hierarchy problem, an
understanding of the observed fermion hierarchies by extra-dimensional
localization effects, and a natural suppression of dangerous 
flavor changing
processes.  A generic feature of these scenarios is that the third
generation is quite different from the first two: by being closer to
the IR brane the top quark avoids an exponential suppression in its
mass, in contrast with the first two generation quarks and leptons,
and can naturally be as heavy as the EW scale.  One can then
generically expect important deviations from the SM in this sector,
most notably anomalous couplings to the $Z$ gauge boson, and
contributions to the Peskin-Takeuchi $T$ parameter.  The first
two generations can also give important contributions to the $S$
parameter.  These constraints tend to put the gauge boson KK
excitations beyond the reach of the LHC, unless these states are
lighter and more weakly coupled than expected, for example due to the
presence of moderately large IR brane kinetic terms.

In the absence of brane localized terms, an $SU(2)_{L} \times
SU(2)_{R}$ bulk gauge symmetry together with a discrete symmetry
exchanging $L$ with $R$ \cite{Agashe:2006at} seem to be essential in
bringing the contributions to the $T$ parameter, and the anomalous
contributions to the $Z \bar{b}_{L} b_{L}$ vertex, under control.
These symmetries, being broken non-locally by boundary conditions,
imply that such effects are calculable.  We have seen that they can
still place important constraints on these models.  A detailed
calculation shows that the 1-loop contributions to the $T$ parameter
are in general sizable.  Furthermore, we have shown that the existence
of $SU(2)_{L} \times SU(2)_{R}$ bidoublets, an essential ingredient in
these scenarios, gives a negative contribution to the $T$ parameter.
Contributions coming from singlets or triplets can compensate such
effects, but the conspiracy occurs in rather well defined regions of
parameter space.  The $S$-$T$ constraints, in particular, very
directly constrain the location of the third generation multiplets.
In this regions, however, KK gauge excitations as light as $3~{\rm
TeV}$ are allowed, thus providing the first example of RS scenarios
with negligible brane localized terms, that are consistent with all EW
precision data and with gauge boson KK states accessible at the LHC.

We also studied the EW constraints in models in which, in addition to the
above structure, the Higgs arises from an extra-dimensional gauge
field, thus alleviating the little hierarchy problem associated with a 5D
scalar field.  These scenarios contain further theoretical relations
due to the embedding of the field content into larger gauge
multiplets, as well as due to the fact that the Yukawa couplings are
related to SM gauge couplings.  In spite of these restrictions, we
find that the EW measurements set bounds on the gauge boson KK masses
of the order of 3~TeV,
similar to those found in the absence of the gauge-Higgs unification
assumption.  An important difference, however, is that the required
multiplet structure, together with the constraints on the location of
the third family, typically lead to light vector-like fermionic
excitations.  Quite generically, a light $SU(2)_{L}$ singlet, that
mixes with the top quark, is expected.  Furthermore, it is likely that
additional light $SU(2)_{L}$ doublet states, some of them with exotic
charges, are present.  All these states can easily have masses in the
$500-800~{\rm GeV}$ range, and should be observable at the LHC, both
by direct (single or pair) production and by their effect through
mixing on the anomalous couplings of the top. Further
fermionic excitations starting as low as $1-2~{\rm TeV}$ are also
generically expected.  All of these should provide interesting
discovery signals and warrant a more detailed study of the associated
phenomenology \cite{gaugeHiggs}.

~\\
~\\  
{\Large \bf Acknowledgements}\\  
~\\  
We would like to thank T.~Tait for valuable conversations and
collaboration at early stages of this work. We would also like to
thank K.~Agashe, J.~Hubisz, B.~Lillie
and A.~Pomarol 
for useful discussions and comments. 
Work at ANL is supported in part by the US DOE, Div.\  
of HEP, Contract W-31-109-ENG-38.  Fermilab is operated by  
Universities Research Association Inc. under contract no.  
DE-AC02-76CH03000  with the DOE. E.P. was supported by DOE under
contract DE-FG02-92ER-40699.
%  
%\clearpage

%-------------------------------------------------------------------------------------------


\begin{thebibliography}{99}  
  
% RS original:  
  
%\cite{Randall:1999ee}  
\bibitem{Randall:1999ee}  
L.~Randall and R.~Sundrum,  
%``A large mass hierarchy from a small extra dimension,''  
Phys.\ Rev.\ Lett.\  {\bf 83}, 3370 (1999)  
[arXiv:hep-ph/9905221].  

%\cite{Peskin:1991sw}  
\bibitem{Peskin:1991sw}  
M.~E.~Peskin and T.~Takeuchi,  
%``Estimation of oblique electroweak corrections,''  
Phys.\ Rev.\ D {\bf 46}, 381 (1992).  
%%CITATION = PHRVA,D46,381;%%  


%\cite{Huber:2000fh}
\bibitem{Huber:2000fh}
  S.~J.~Huber and Q.~Shafi,
  %``Higgs mechanism and bulk gauge boson masses in the Randall-Sundrum
  %model,''
  Phys.\ Rev.\ D {\bf 63}, 045010 (2001)
  [arXiv:hep-ph/0005286];
  %%CITATION = HEP-PH 0005286;%%
%\cite{Huber:2001gw}
%\bibitem{Huber:2001gw}
  S.~J.~Huber, C.~A.~Lee and Q.~Shafi,
  %``Kaluza-Klein excitations of W and Z at the LHC?,''
  Phys.\ Lett.\ B {\bf 531}, 112 (2002)
  [arXiv:hep-ph/0111465];
  %%CITATION = HEP-PH 0111465;%%
%\cite{Csaki:2002gy}
%\bibitem{Csaki:2002gy}
  C.~Csaki, J.~Erlich and J.~Terning,
  %``The effective Lagrangian in the Randall-Sundrum model and electroweak
  %physics,''
  Phys.\ Rev.\ D {\bf 66}, 064021 (2002)
  [arXiv:hep-ph/0203034];
  %%CITATION = HEP-PH 0203034;%%
%\cite{Hewett:2002fe}
%\bibitem{Hewett:2002fe}
  J.~L.~Hewett, F.~J.~Petriello and T.~G.~Rizzo,
  %``Precision measurements and fermion geography in the Randall-Sundrum  model
  %revisited,''
  JHEP {\bf 0209}, 030 (2002)
  [arXiv:hep-ph/0203091].
  %%CITATION = HEP-PH 0203091;%%


%\cite{Carena:2002dz}
\bibitem{Carena:2002dz}
  M.~Carena, E.~Pont\'{o}n, T.~M.~P.~Tait and C.~E.~M.~Wagner,
  %``Opaque branes in warped backgrounds,''
  Phys.\ Rev.\ D {\bf 67}, 096006 (2003)
  [arXiv:hep-ph/0212307];
  %%CITATION = HEP-PH 0212307;%%
%\cite{Davoudiasl:2002ua}
%\bibitem{Davoudiasl:2002ua}
  H.~Davoudiasl, J.~L.~Hewett and T.~G.~Rizzo,
  %``Brane localized kinetic terms in the Randall-Sundrum model,''
  Phys.\ Rev.\ D {\bf 68}, 045002 (2003)
  [arXiv:hep-ph/0212279].
  %%CITATION = HEP-PH 0212279;%%


%\cite{Carena:2003fx}
\bibitem{Carena:2003fx}
  M.~Carena, A.~Delgado, E.~Pont\'{o}n, T.~M.~P.~Tait and C.~E.~M.~Wagner,
  %``Precision electroweak data and unification of couplings in warped extra
  %dimensions,''
  Phys.\ Rev.\ D {\bf 68}, 035010 (2003)
  [arXiv:hep-ph/0305188].
  %%CITATION = HEP-PH 0305188;%%

%
\bibitem{Carena:2004zn}
  M.~Carena, A.~Delgado, E.~Pont\'{o}n, T.~M.~P.~Tait and C.~E.~M.~Wagner,
  %``Warped fermions and precision tests,''
  Phys.\ Rev.\ D {\bf 71}, 015010 (2005)
  [arXiv:hep-ph/0410344].
  %%CITATION = HEP-PH 0410344;%%


%\cite{Agashe:2003zs}
\bibitem{Agashe:2003zs}
K.~Agashe, A.~Delgado, M.~J.~May and R.~Sundrum,
%``RS1, custodial isospin and precision tests,''
JHEP {\bf 0308}, 050 (2003)
[arXiv:hep-ph/0308036].
%%CITATION = HEP-PH 0308036;%%


%\cite{Agashe:2006at}
\bibitem{Agashe:2006at}
  K.~Agashe, R.~Contino, L.~Da Rold and A.~Pomarol,
  %``A custodial symmetry for Z b anti-b,''
  arXiv:hep-ph/0605341.
  %%CITATION = HEP-PH 0605341;%%

\bibitem{gauge:Higgs:unification}
%\cite{Manton:1979kb}
%\bibitem{Manton:1979kb}
  N.~S.~Manton,
  %``A New Six-Dimensional Approach To The Weinberg-Salam Model,''
  Nucl.\ Phys.\ B {\bf 158}, 141 (1979);
  %%CITATION = NUPHA,B158,141;%%
%\cite{Hosotani:1983xw}
%\bibitem{Hosotani:1983xw}
  Y.~Hosotani,
  %``Dynamical Mass Generation By Compact Extra Dimensions,''
  Phys.\ Lett.\ B {\bf 126}, 309 (1983);
  %%CITATION = PHLTA,B126,309;%%
%\cite{Hatanaka:1998yp}
%\bibitem{Hatanaka:1998yp}
  H.~Hatanaka, T.~Inami and C.~S.~Lim,
 %  ``The gauge hierarchy problem and higher dimensional gauge theories,''
  %
  Mod.\ Phys.\ Lett.\ A {\bf 13}, 2601 (1998)
  [arXiv:hep-th/9805067];
  %%CITATION = HEP-TH 9805067;%%
%\cite{Antoniadis:2001cv}
%\bibitem{Antoniadis:2001cv}
  I.~Antoniadis, K.~Benakli and M.~Quiros,
  %``Finite Higgs mass without supersymmetry,''
  New J.\ Phys.\  {\bf 3}, 20 (2001)
  [arXiv:hep-th/0108005];
  %%CITATION = HEP-TH 0108005;%%
%\cite{Kubo:2001zc}
%\bibitem{Kubo:2001zc}
  M.~Kubo, C.~S.~Lim and H.~Yamashita,
%   ``The Hosotani mechanism in bulk gauge theories with an orbifold
%  extra  space 
%   S(1)/Z(2),'' 
  %
  Mod.\ Phys.\ Lett.\ A {\bf 17}, 2249 (2002)
  [arXiv:hep-ph/0111327];
  %%CITATION = HEP-PH 0111327;%%
%\cite{vonGersdorff:2002as}
%\bibitem{vonGersdorff:2002as}
  G.~von Gersdorff, N.~Irges and M.~Quiros,
  %``Bulk and brane radiative effects in gauge theories on orbifolds,''
  Nucl.\ Phys.\ B {\bf 635}, 127 (2002)
  [arXiv:hep-th/0204223];
  %%CITATION = HEP-TH 0204223;%%
%\cite{Csaki:2002ur}
%\bibitem{Csaki:2002ur}
  C.~Csaki, C.~Grojean and H.~Murayama,
  %``Standard model Higgs from higher dimensional gauge fields,''
  Phys.\ Rev.\ D {\bf 67}, 085012 (2003)
  [arXiv:hep-ph/0210133];
  %%CITATION = HEP-PH 0210133;%%
%\cite{Haba:2002py}
%\bibitem{Haba:2002py}
  N.~Haba, M.~Harada, Y.~Hosotani and Y.~Kawamura,
  %``Dynamical rearrangement of gauge symmetry on the orbifold S(1)/Z(2),''
  Nucl.\ Phys.\ B {\bf 657}, 169 (2003)
  [Erratum-ibid.\ B {\bf 669}, 381 (2003)]
  [arXiv:hep-ph/0212035];
  %%CITATION = HEP-PH 0212035;%%
%\cite{Scrucca:2003ra}
%\bibitem{Scrucca:2003ra}
  C.~A.~Scrucca, M.~Serone and L.~Silvestrini,
  %``Electroweak symmetry breaking and fermion masses from extra dimensions,''
  Nucl.\ Phys.\ B {\bf 669}, 128 (2003)
  [arXiv:hep-ph/0304220];
  %%CITATION = HEP-PH 0304220;%%
%\cite{Scrucca:2003ut}
%\bibitem{Scrucca:2003ut}
  C.~A.~Scrucca, M.~Serone, L.~Silvestrini and A.~Wulzer,
  %``Gauge-Higgs unification in orbifold models,''
  JHEP {\bf 0402}, 049 (2004)
  [arXiv:hep-th/0312267];
  %%CITATION = HEP-TH 0312267;%%
%\cite{Haba:2004qf}
%\bibitem{Haba:2004qf}
  N.~Haba, Y.~Hosotani, Y.~Kawamura and T.~Yamashita,
%   ``Dynamical symmetry breaking in gauge-Higgs unification on orbifold,''
  %
  Phys.\ Rev.\ D {\bf 70}, 015010 (2004)
  [arXiv:hep-ph/0401183];
  %%CITATION = HEP-PH 0401183;%%
%\cite{Biggio:2004kr}
%\bibitem{Biggio:2004kr}
  C.~Biggio and M.~Quiros,
  %``Higgs-gauge unification without tadpoles,''
  Nucl.\ Phys.\ B {\bf 703}, 199 (2004)
  [arXiv:hep-ph/0407348];
  %%CITATION = HEP-PH 0407348;%%
%\cite{Hosotani:2004wv}
%\bibitem{Hosotani:2004wv}
  Y.~Hosotani, S.~Noda and K.~Takenaga,
%   ``Dynamical gauge-Higgs unification in the electroweak theory,''
  %
  Phys.\ Lett.\ B {\bf 607}, 276 (2005)
  [arXiv:hep-ph/0410193];
  %%CITATION = HEP-PH 0410193;%%
%\cite{Cacciapaglia:2005da}
%\bibitem{Cacciapaglia:2005da}
  G.~Cacciapaglia, C.~Csaki and S.~C.~Park,
%   ``Fully radiative electroweak symmetry breaking,''
  %
  JHEP {\bf 0603}, 099 (2006)
  [arXiv:hep-ph/0510366];
  %%CITATION = HEP-PH 0510366;%%
%\cite{Panico:2005dh}
%\bibitem{Panico:2005dh}
  G.~Panico, M.~Serone and A.~Wulzer,
  %``A model of electroweak symmetry breaking from a fifth dimension,''
  Nucl.\ Phys.\ B {\bf 739}, 186 (2006)
  [arXiv:hep-ph/0510373];
  %%CITATION = HEP-PH 0510373;%%
%\cite{Panico:2006em}
%\bibitem{Panico:2006em}
  G.~Panico, M.~Serone and A.~Wulzer,
%   ``Electroweak symmetry breaking and precision tests with a fifth
%  dimension,'' 
  %
  arXiv:hep-ph/0605292.
  %%CITATION = HEP-PH 0605292;%%


%\cite{Agashe:2004rs}
\bibitem{Agashe:2004rs}
  K.~Agashe, R.~Contino and A.~Pomarol,
  %``The minimal composite Higgs model,''
  Nucl.\ Phys.\ B {\bf 719}, 165 (2005)
  [arXiv:hep-ph/0412089];
  %%CITATION = HEP-PH 0412089;%%




%\cite{Huber:2002sp}
\bibitem{Huber:2002sp}
  S.~J.~Huber,
  %``Flavor physics and warped extra dimensions,''
  arXiv:hep-ph/0211056;
  %%CITATION = HEP-PH 0211056;%%%\cite{Huber:2003tu}
%\bibitem{Huber:2003tu}
  S.~J.~Huber,
  %``Flavor violation and warped geometry,''
  Nucl.\ Phys.\ B {\bf 666} (2003) 269
  [arXiv:hep-ph/0303183].
  %%CITATION = HEP-PH 0303183;%%
%\cite{Agashe:2004ay}  
%\bibitem{Agashe:2004ay}  
K.~Agashe, G.~Perez and A.~Soni,  
%``B-factory signals for a warped extra dimension,''  
arXiv:hep-ph/0406101; 
%%CITATION = HEP-PH 0406101;%%  
%\cite{Agashe:2004cp}
%\bibitem{Agashe:2004cp}
%K.~Agashe, G.~Perez and A.~Soni,
%``Flavor structure of warped extra dimension models,''
arXiv:hep-ph/0408134.
%%CITATION = HEP-PH 0408134;%%

\bibitem{gaugeHiggs}
  M.~Carena, E.~Pont\'{o}n, J.~Santiago and C.~E.~M.~Wagner, in
  preparation 


%\cite{Haber:1999zh}
\bibitem{Haber:1999zh}
  H.~E.~Haber and H.~E.~Logan,
%   ``Radiative corrections to the Z b anti-b vertex 
%and constraints on  extended
%   Higgs sectors,''
  %
  Phys.\ Rev.\ D {\bf 62}, 015011 (2000)
  [arXiv:hep-ph/9909335].
  %%CITATION = HEP-PH 9909335;%%

%\cite{Choudhury:2001hs}
\bibitem{Beauty}
  D.~Choudhury, T.~M.~P.~Tait and C.~E.~M.~Wagner,
  %``Beautiful mirrors and precision electroweak data,''
  Phys.\ Rev.\ D {\bf 65}, 053002 (2002)
  [arXiv:hep-ph/0109097].
  %%CITATION = HEP-PH 0109097;%%


%\cite{Agashe:2005dk}
\bibitem{Agashe:2005dk}
  K.~Agashe and R.~Contino,
  %``The minimal composite Higgs model and electroweak precision tests,''
  Nucl.\ Phys.\ B {\bf 742}, 59 (2006)
  [arXiv:hep-ph/0510164].
  %%CITATION = HEP-PH 0510164;%%

%\cite{Lavoura:1992np}
\bibitem{Lavoura:1992np}
  L.~Lavoura and J.~P.~Silva,
  %``The Oblique corrections from vector - like singlet and doublet quarks,''
  Phys.\ Rev.\ D {\bf 47}, 2046 (1993).
  %%CITATION = PHRVA,D47,2046;%%

%\cite{Davoudiasl:2005uu}
\bibitem{Davoudiasl:2005uu}
  H.~Davoudiasl, B.~Lillie and T.~G.~Rizzo,
  %``Off-the-wall Higgs in the universal Randall-Sundrum model,''
  arXiv:hep-ph/0508279.
  %%CITATION = HEP-PH 0508279;%%

%\cite{Agashe:2004bm}
\bibitem{Agashe:2004bm}
  K.~Agashe and G.~Servant,
   ``Baryon number in warped GUTs: Model building and (dark matter related)
  %phenomenology,''
  JCAP {\bf 0502}, 002 (2005)
  [arXiv:hep-ph/0411254].
  %%CITATION = HEP-PH 0411254;%%

%\cite{Hirn:2006nt}
\bibitem{Hirn:2006nt}
  J.~Hirn and V.~Sanz,
%   ``A negative S parameter from holographic technicolor,''
  %
  arXiv:hep-ph/0606086.
  %%CITATION = HEP-PH 0606086;%%
  

%\cite{unknown:2005em}
\bibitem{unknown:2005em}
    [ALEPH Collaboration],
%   ``Precision electroweak measurements on the Z resonance,''
  %
  Phys.\ Rept.\  {\bf 427}, 257 (2006)
  [arXiv:hep-ex/0509008].
  %%CITATION = HEP-EX 0509008;%%




%\cite{delAguila:2000rc}
\bibitem{delAguila:2000rc}
  F.~del Aguila, M.~Perez-Victoria and J.~Santiago,
  %``Observable contributions of new exotic quarks to quark mixing,''
  JHEP {\bf 0009}, 011 (2000)
  [arXiv:hep-ph/0007316];
  %%CITATION = HEP-PH 0007316;%%
%\cite{DelAguila:2001pu}
%\bibitem{DelAguila:2001pu}
F.~del Aguila and J.~Santiago,
%   ``Signals from extra dimensions decoupled from the compactification
%  scale,'' 
JHEP {\bf 0203}, 010 (2002)
[arXiv:hep-ph/0111047];
%%CITATION = HEP-PH 0111047;%%
%\cite{Aguilar-Saavedra:2002kr}
%\bibitem{Aguilar-Saavedra:2002kr}
  J.~A.~Aguilar-Saavedra,
  %``Effects of mixing with quark singlets,''
  Phys.\ Rev.\ D {\bf 67}, 035003 (2003)
  [Erratum-ibid.\ D {\bf 69}, 099901 (2004)]
  [arXiv:hep-ph/0210112].
  %%CITATION = HEP-PH 0210112;%%

\bibitem{Amidei:1996dt}
  D.~Amidei {\it et al.}  [TeV-2000 Study Group],
%   ``Future electroweak physics at the Fermilab Tevatron: Report of the
%   TeV-2000 Study Group,''
  %
SLAC-REPRINT-1996-085 



%\cite{Altarelli:2000ye}
\bibitem{Altarelli:2000ye}
  G.~Altarelli and M.~L.~Mangano,
  %``Standard model physics (and more) at the LHC.  Proceedings, Workshop,
  %Geneva, Switzerland, May 25-26, October 14-15,  1999,''
CERN-2000-004
%\href{http://www.slac.stanford.edu/spires/find/hep/www?r=cern-2000-004}{SPIRES entry}


%\cite{Agashe:2006wa}
\bibitem{Agashe:2006wa}
  K.~Agashe, G.~Perez and A.~Soni,
  %``Collider Signals of Top Quark Flavor Violation from a Warped Extra
  %Dimension,''
  arXiv:hep-ph/0606293.
  %%CITATION = HEP-PH 0606293;%%



%\cite{Andre:2003wc}
\bibitem{Rosner}
  T.~C.~Andre and J.~L.~Rosner,
%   ``Exotic Q = -1/3 quark signatures at high-energy hadron colliders,''
  %
  Phys.\ Rev.\ D {\bf 69}, 035009 (2004)
  [arXiv:hep-ph/0309254].
  %%CITATION = HEP-PH 0309254;%%
%\cite{Morrissey:2003sc}
\bibitem{Morrissey:2003sc}
  D.~E.~Morrissey and C.~E.~M.~Wagner,
 %  ``Beautiful mirrors, unification of couplings and collider phenomenology,''
  %
  Phys.\ Rev.\ D {\bf 69}, 053001 (2004)
  [arXiv:hep-ph/0308001].
  %%CITATION = HEP-PH 0308001;%%


%\cite{Aguilar-Saavedra:2005pv}
\bibitem{Aguilar-Saavedra:2005pv}
  J.~A.~Aguilar-Saavedra,
  %``Pair production of heavy Q = 2/3 singlets at LHC,''
  Phys.\ Lett.\ B {\bf 625}, 234 (2005)
  [Erratum-ibid.\ B {\bf 633}, 792 (2006)]
  [arXiv:hep-ph/0506187];
  %%CITATION = HEP-PH 0506187;%%
%\cite{Aguilar-Saavedra:2006gv}
%\bibitem{Aguilar-Saavedra:2006gv}
  J.~A.~Aguilar-Saavedra,
  %``New signals in pair production of heavy Q = 2/3 singlets at LHC,''
  PoS {\bf TOP2006}, 003 (2006)
  [arXiv:hep-ph/0603199].
  %%CITATION = HEP-PH 0603199;%%

\bibitem{T:littleHiggs}
D.~Costanzo, ATL-PHYS-2004-004;
%\cite{Azuelos:2004dm}
%\bibitem{Azuelos:2004dm}
  G.~Azuelos {\it et al.},
  %``Exploring little Higgs models with ATLAS at the LHC,''
  Eur.\ Phys.\ J.\ C {\bf 39S2}, 13 (2005)
  [arXiv:hep-ph/0402037].
  %%CITATION = HEP-PH 0402037;%%

%\cite{Bamert:1996px}
\bibitem{Bamert:1996px}
  P.~Bamert, C.~P.~Burgess, J.~M.~Cline, D.~London and E.~Nardi,
  %``R_b and New Physics: A Comprehensive Analysis,''
  Phys.\ Rev.\ D {\bf 54}, 4275 (1996)
  [arXiv:hep-ph/9602438].
  %%CITATION = HEP-PH 9602438;%%


\end{thebibliography}
\end{document}